\documentclass[%
aip,
jcp,%
amsmath,amssymb,
reprint,%
noeprint,
nopmid,
]{revtex4-2}
\usepackage[latin1]{inputenc}

\usepackage{amsmath}%
\usepackage{amsfonts}
\usepackage{amssymb,bm}
\usepackage{graphicx}
\usepackage{physics}
\usepackage{upgreek}
\usepackage{booktabs}
\usepackage[outercaption]{sidecap}   
\usepackage{color}
\usepackage[normalem]{ulem}
\mathchardef\mhyphen="2D
\usepackage{txfonts}
\usepackage{lipsum}
\usepackage{comment}
\usepackage{esvect}

\usepackage{bm}
\renewcommand{\vec}{\bm}

\newcommand{\cff}{c_{\mathrm{ff}}}
\newcommand{\e}{\mathrm{e}}
\newcommand{\ii}{\mathrm{i}}
\newcommand{\Zr}{Z_\mathrm{r}}
\newcommand{\betac}{\beta_\mathrm{c}}
\newcommand{\Tc}{T_\mathrm{c}}

\newcommand{\Gammax}{\Gamma_1^{(x)}}
\newcommand{\Gammat}{\Gamma_1^{(t)}}
\newcommand{\GammaZr}{\Gamma_1^{(\mathrm{r})}}

\usepackage[version=4]{mhchem}
\usepackage{xspace}

\newcommand{\HHtwo}{\ce{H + H2}\xspace}
\newcommand{\DDtwo}{\ce{D + D2}\xspace}
\newcommand{\TTtwo}{\ce{T + T2}\xspace}

\newcommand{\pia}{p'_{a}}
\newcommand{\pfa}{p''_{a}}
\newcommand{\pib}{p'_{b}}
\newcommand{\pfb}{p''_{b}}

\newcommand{\dta}{\delta \tau_a}
\newcommand{\dtb}{\delta \tau_b}

\newcommand{\tot}{\tau^{\mathrm{tot}}}

\usepackage{dcolumn} %
\newcolumntype{d}[1]{D{.}{.}{#1}} %
\newcommand{\head}[1]{\multicolumn{1}{c}{#1}}

\usepackage{hyperref}

\renewcommand{\d}{\mathrm{d}}

\begin{document}

\def\bra#1{\left<{#1}\right|}
\def\ket#1{\left|{#1}\right>}

\title{Perturbatively corrected ring-polymer instanton rate theory rigorously captures anharmonicity and deep tunneling}
\author{Jindra Du\v sek}
\affiliation{Department of Chemistry and Applied Biosciences, ETH Z\"urich, 8093 Z\"urich, Switzerland}
\author{Joseph E. Lawrence}
\affiliation{Department of Chemistry and Applied Biosciences, ETH Z\"urich, 8093 Z\"urich, Switzerland}
\affiliation{\mbox{Simons Center for Computational Physical Chemistry, New York University, New York, NY 10003, USA}}
\affiliation{Department of Chemistry, New York University, New York, NY 10003, USA}
\author{Jeremy O. Richardson}
\email{jeremy.richardson@phys.chem.ethz.ch}
\affiliation{Department of Chemistry and Applied Biosciences, ETH Z\"urich, 8093 Z\"urich, Switzerland}

\begin{abstract}
In this paper, we derive a perturbatively-corrected instanton rate theory in the ring-polymer framework (RPI+PC), which significantly enhances the accuracy of instanton theory by using third and fourth derivatives of the potential to capture anharmonic effects.
Instanton theory is a rigorous semiclassical method that extends transition-state theory by including quantum tunneling along a well-defined optimal tunneling pathway.
However, the standard leading-order instanton theory (RPI) neglects anharmonicity perpendicular to this tunneling path.
The RPI+PC method described here corrects this using only local information along the same instanton trajectory as the leading-order theory. 
Hence, RPI+PC does not require a global potential energy surface and is readily applicable in combination with ab initio electronic-structure methods.
The derivation of the RPI+PC result is performed within the flux-correlation formalism using standard techniques from asymptotic analysis, and the final rate expression is shown to be independent of the choice of dividing surface.
We demonstrate that RPI+PC represents a systematic improvement over RPI by analyzing its asymptotic properties in the semiclassical limit ($\hbar\to0$ with total thermal time $\tot=\beta\hbar$ kept constant) and illustrate its improved performance on a series of model systems for which exact results are available for comparison, including the collinear $\HHtwo$ reaction and its isotopic variants.
\end{abstract}
\maketitle

\section{Introduction}

Accurately predicting the rate of chemical reactions is central to %
computational chemistry.
Such predictions are important across a wide range of fields, including atmospheric chemistry, astrochemistry and catalysis.
Historically, computational chemistry has failed to live up to the ambition of being fully predictive, and it is still common to rely on experimental data for the rates of elementary reactions when modeling complex reaction networks.
However, with growing computational power along with recent advances in machine learning, we are entering an era in which completely predictive modeling is becoming feasible.

To reliably predict reaction rates, both high-quality electronic-structure calculations, as well as accurate rate theories will be required. The present work focuses on developing the latter.
When one searches for high accuracy, it is perhaps natural to consider exact wavefunction methods, such as quantum reactive scattering.\cite{RecentAdvancesinQRS} However, the exponential scaling of quantum mechanics means these techniques are still fundamentally limited to systems containing only a handful of atoms.
Furthermore, they require full potential energy surfaces (PESs) and expert knowledge to apply.

For this reason the most commonly used technique for predicting reaction rates is still Eyring transition state theory (TST)\cite{Eyring1935} with rigid-rotor and harmonic-oscillator (RRHO) approximations for the rotations and vibrations.
While this may give reasonable results for certain simple systems, it is limited by its failure to describe anharmonicity or tunneling. In particular, tunneling is known to play an important role at low temperature, especially for hydrogen-atom transfers,\cite{BellBook} and anharmonic effects can have a strong influence on the rate, 
particularly when low-frequency bending modes or torsions are coupled to the reaction coordinate.\cite{TruhlarTorsionReaction,TorsionalAnharmRSC,AnharmMD}

\emph{Semiclassical transition-state theory} (SCTST)\cite{Miller1990SCTST,Nguyen2010SCTST,NewDevelopmentsSCTST} was developed to go beyond the RRHO approximations of Eyring TST\@.
In its standard implementation, it utilizes the framework of vibrational perturbation theory (VPT2) to account for anharmonicity (at least approximately) based on the third and fourth derivatives of the potential energy surface at the barrier top.
However, while it can capture shallow-tunneling effects, it is known to break down in the deep-tunneling regime\cite{Wagner2013SCTST,Goel2018SCTST} as it only contains local information at the saddle point.
\footnote{Although numerous corrections have been proposed to improve the behavior of SCTST in the deep-tunneling regime, they rely on a predefined ansatz of the tunneling behavior which cannot be expected to behave correctly in all cases.}

For systems exhibiting deep tunneling, an alternative semiclassical approach known as \emph{instanton theory}\cite{Miller1975semiclassical,Perspective} is preferred.
Instanton theory describes deep tunneling in a rigorous manner through an asymptotic approximation which becomes exact in the $\hbar\rightarrow0$ limit.
Unlike TST and SCTST, which rely on information from the saddle-point geometry only, instanton theory is based on an optimal tunneling pathway called the \emph{instanton trajectory}. This trajectory differs from the minimal-energy pathway, in a phenomenon known as \emph{corner-cutting},\cite{Chapman1975rates} enabling an accurate description of deep tunneling in multidimensional systems.
However, instanton theory treats the vibrational modes perpendicular to the tunneling pathway within a harmonic approximation.
Importantly, however, instanton theory is based on rigorous asymptotic analysis, and is therefore systematically improvable by considering higher-order terms in the \emph{asymptotic series}.

In this work, we extend the asymptotic approximation of instanton theory to the next order in this asymptotic series, and in this way capture anharmonic effects in a perturbative manner.
We work within the framework of ring-polymer instanton theory (RPI)\cite{InstReview}  in order to derive a practical approach for the calculation of this perturbative correction (PC) in chemical systems.
The resulting theory (RPI+PC) combines the advantages of both SCTST and instanton theory to provide a rigorous description of deep tunneling with anharmonic effects.

The derivation and implementation of the new theory has many similarities to our recent perturbative correction to the instanton theory of tunneling splittings.\cite{PCIT2023} The key difference is that calculating the rate involves an additional integral over time.
In both cases, the RPI+PC theory can be implemented as a post-processing step after a standard ring-polymer instanton optimization has been carried out.
In addition to the potentials, gradients and Hessians along the instanton pathway, the new method also requires third and fourth derivatives in order to account for anharmonicity---reminiscent of SCTST.

There are of course other techniques which capture anharmonicity and tunneling, most notably ring-polymer molecular dynamics (RPMD) rate theory,\cite{RPMDrate,RPMDrefinedRate,Habershon2013RPMDreview} which provides a complementary approach to instanton theory. 
In fact, the justification for using RPMD in the deep-tunneling regime stems from its connection to semiclassical instanton theory.\cite{RPInst}
A key advantage of instanton theory is that it only uses \emph{local} information along the instanton to calculate the rate, whereas RPMD requires sampling over the entire \emph{global} potential energy surface.
Of course, a major advantage of RPMD is in studying reactions in liquids, where the large number of tunneling paths means that the steepest-descent approximation of instanton theory cannot be applied.
However, outside of this particular limitation, instanton theory's rigorous first-principle derivation provides several advantages.
In particular, where the assumptions of the standard theory break down
(such as for broad-topped barriers, in microcanonical ensembles and for reactions with prereactive complexes) %
its rigorous derivation has enabled it to be extended to treat these problems.\cite{broadtop,Faraday,DoSTMI,JoeFaraday,RPMDbreakdown}
An additional advantage of the present approach over RPMD is that it not only describes anharmonic fluctuations perpendicular to the path, but also \emph{along} the path. 
Although RPMD gives an exact description of the quantum statistics at the transition state, as pointed out by Pollak and Cao,\cite{Pollak2022hbarSquared_Correction}  its description of tunneling does not incorporate the higher order semiclassical corrections that describe anharmonic fluctuations along the tunneling path. As discussed in the Appendix, the first order RPI+PC correction developed in this work is consistent with the ``$\hbar^2$'' correction of Pollak and co-workers.\cite{Pollak2022hbarSquared_Correction,PollakRatehbar2} 

In its original form, instanton theory is only applicable for temperatures below the crossover temperature, $T_{\mathrm{c}}$, where the instanton collapses. 
There have been many approaches suggested to fix this problem,\cite{Haenggi1988crossover,Cao1996QTST,Kryvohuz2011rate,Zhang2014interpolation,Faraday,McConnell2017instanton,PollakRateUniform}
including, most recently, the generalization to a uniform asymptotic series.\cite{JoeUniform}
Since our first-order perturbative correction is derived within the framework of rigorous asymptotics, in principle it can be combined with this uniform asymptotic approach to obtain the first-order corrected instanton rate at arbitrary temperatures. This will be the subject of upcoming work.

The paper is structured as follows. Section \ref{sec:Overview_theory} gives an overview of the theory and its derivation, explaining the key concepts and structure of the derivation. Section~\ref{sec:cff} then covers the details of the derivation in the RPI framework along with the key working equations. Section~\ref{sec:results}  then applies the new RPI+PC theory to a series of one-dimensional and multidimensional benchmarks, illustrating the accuracy of the method. Finally, Sec.~\ref{sec:conclusion} concludes and discusses the outlook for future work.

\section{Theoretical Overview}
\label{sec:Overview_theory}

The focus of this work is the calculation of the first-order correction to the thermal rate constant using a rigorous asymptotic expansion in $\hbar$.
It is important to clarify exactly what we mean by this at the outset.
The rate constant is usually written as a function of temperature, $T$, or equivalently of inverse temperature $\beta = 1/(k_{\mathrm{B}}T)$, where $k_{\mathrm{B}}$ is the Boltzmann constant.
It may, therefore, seem natural to consider the asymptotic behavior as $\hbar\to0$ while keeping $\beta$ or $T$ constant. 
However, in the derivation of semiclassical methods, such as instanton theory, it is the thermal time $\tot = \beta \hbar$ that is kept constant.
This keeps the semiclassical analysis of real-time effects on an equal footing with thermal effects, by simply reinterpreting the Boltzmann operator as an ``imaginary-time'' propagator, $\mathrm{e}^{-\beta \hat{H}}=\mathrm{e}^{-\tot \hat{H}/\hbar}$.   
It turns out that this results in an expansion around a \emph{semiclassical}---rather than a classical---limit, where crucially the leading-order term already captures quantum effects such as tunneling and zero-point energy. %

The starting point for our derivation is the quantum-mechanical definition of the thermal rate in terms of the \emph{flux-correlation function} formalism,\cite{Miller1983rate}
\begin{equation}
\label{eq:k}
    k  = \frac{1}{2\Zr} \int_{-\infty}^\infty  \cff(t) \: \mathrm{d}t
    \,,
\end{equation}
where $\Zr$ %
is the reactant partition function %
and $\cff(t)$ is the flux--flux correlation function
\begin{equation}
\label{eq:cff}
\cff(t) =
\tr \left[
\e^{-\ii\hat{H}t/\hbar}
\,\hat{F}
\,\e^{-\hat{H}(\tot-\ii t)/\hbar}
\,\hat{F}
\right]
\,.
\end{equation}
Here, $\hat{H}$ is the Hamiltonian and $\hat{F}$ is the \emph{flux operator}, which  measures the flux through a dividing surface defined by $\sigma(x)=0$ and can be written as follows:
\begin{equation}
\label{eq:Flux_def}
\hat{F} =
\frac{1}{2m} \left[
\delta(\sigma(\hat{x}))
\,\hat{p}_\sigma
+
\hat{p}^\dagger_\sigma
\,\delta(\sigma(\hat{x}))
\right]
\end{equation}
with $\hat{p}_\sigma=\frac{\partial \sigma}{\partial x} \cdot \hat{p}$ being the momentum normal to the surface.
Importantly, while the theory uses the concept of a dividing surface, the rate is formally independent of how we choose it. Note that throughout, without loss of generality, we work with the convention that the ($f$-dimensional) coordinate vector, $x$, has been scaled such that each coordinate has the same mass, $m$.

Due to the exponential scaling of quantum mechanics, it is impractical to evaluate $\cff(t)$ exactly for complex molecular systems containing many degrees of freedom.
This is where asymptotic methods, such as instanton theory, are useful. Asymptotic analysis effectively extends perturbation theory beyond simple Taylor expansions.
As with perturbation theory, its power is that it simplifies complex problems into a series of systematically improvable approximations. 
The particular kind of asymptotic series we use in the present work is an asymptotic power series, 
such that the rate constant may be expanded as
\begin{subequations}\label{eq:asymptotic_series}
\begin{equation}
    k(\hbar)  \sim k_0(\hbar)\left[ 1+  \sum_{n=1}^\infty \hbar^n \Gamma_n  \right],
\end{equation}
as $\hbar\to 0$. Here the symbol ``$\sim$'' is a formal symbol from asymptotic analysis, which should be read ``is asymptotically equal to'' and in our context it has the following meaning
\begin{equation}
    \lim_{\hbar\to 0}\frac{ k(\hbar)}{k_0(\hbar)} =1  
\end{equation}
\begin{equation}
    \lim_{\hbar\to 0}\frac{ k(\hbar) - k_0(\hbar)\left[ 1+  \sum_{m=1}^{n-1} \hbar^m \Gamma_m  \right]}{\hbar^n} =\Gamma_n  
\end{equation}
\end{subequations}
for all $n\geq 1$. 
This is why semiclassical methods are rigorous approximations: they become exact in the limit $\hbar\to0$.
For a more detailed introduction to this branch of mathematics, we refer the reader to the classic textbook by Bender and Orszag\cite{BenderBook} and to Appendix B of our previous paper.\cite{PCIT2023}

In obtaining the asymptotic series for the thermal rate, we proceed in three key steps.
First, we will make use of the path-integral formulation of quantum mechanics (in particular using the \emph{ring-polymer instanton} (RPI) framework\cite{InstReview}) to expand $\cff(t)$ as an asymptotic power series of the form
\begin{equation}
\label{eq:cff_expansion}
\cff(t) \sim \sum_{\rm traj.} \frac{A(t)}{\hbar} \,\e^{-S(t)/\hbar} \left[1+\hbar \Gamma_1^{(x)}(t) + \mathcal{O}(\hbar^2)
\right].
\end{equation}
Here the sum is over classical trajectories that correspond to stationary-action paths.
Each trajectory has an associated asymptotic series, with $\frac{A(t)}{\hbar} \,\e^{-S(t)/\hbar}$ forming the \emph{leading-order} term in the asymptotic series and $\Gamma_1^{(x)}$ the \emph{first-order correction}.
Full details about the definitions of these terms and their derivation are given in Sec.~\ref{sec:cff}.

Second, we plug the expanded flux--flux correlation function from Eq.~\eqref{eq:cff_expansion} into the rate formula in Eq.~\eqref{eq:k}. To evaluate the time-integral, one deforms the contour of integration, $\gamma$, into the complex plane such that %
it passes through the \emph{stationary time} $\tilde{t}$, for which $\partial_t S(t)\lvert_{t=\tilde{t}}\,= 0$.
The rate can then be related to the following one-dimensional integral
\begin{align}
\label{eq:Khbar}
kZ_{\mathrm{r}} \sim  \int_\gamma \frac{A(t)}{\hbar} \,\e^{-S(t)/\hbar} \left[1+\hbar \Gamma_1^{(x)}(t) + \mathcal{O}(\hbar^2) \right] 
\d t
 \,.
\end{align}
Here, a careful analysis (confirmed by numerical results) shows that, while there are multiple paths contributing to the asymptotic expression for $\cff(t)$,\cite{AdiabaticGreens,InstReview} only periodic orbits contribute to the rate in Eq.~\eqref{eq:Khbar}; this remains true even beyond the leading-order term, thus eliminating the need for introducing extra projection operators.\cite{QInst}
When the dividing surface is chosen within the barrier region, these paths have stationary times which are purely imaginary, $\tilde{t}=-\ii\tilde{\tau}$, corresponding to a standard purely imaginary-time instanton orbit.\cite{Miller1975semiclassical} It is straightforward to show that there are only two such periodic orbits, which correspond to opposite signs of the imaginary momentum, each of which contribute equally to the rate [and thus considering just one of them cancels the factor of $1/2$ in Eq.~\eqref{eq:k}].

With this in hand we again employ asymptotic methods to evaluate the time-integral in Eq.~\eqref{eq:Khbar} to give
\begin{equation}
\label{eq:kZr_asymptotic}
    kZ_{\mathrm{r}} \sim (k\Zr)_{\rm inst,0} \left[ 1+ \hbar \left(\Gammax + \Gammat\right) + \mathcal{O}(\hbar^2)\right]
    \,,
\end{equation}
where $(k\Zr)_{\rm inst,0}$ is the well-known \emph{leading-order} instanton result.\cite{Miller1975semiclassical,InstReview} Following this derivation we find that the first-order correction can be decomposed into two terms: the \emph{spatial correction}, $\Gammax=\Gammax(\tilde{t})$, which is the correction to the correlation function from Eq.~\eqref{eq:cff_expansion} evaluated at the stationary time, for which full details are given in Sec.~\ref{sec:cff_inst}, and the \emph{temporal correction} $\Gammat$, which arises from the first-order approximation to the integral in Eq.~\eqref{eq:Khbar}, and for which full details are given in Sec.~\ref{sec:t_integration}. 

The third and final ingredient needed to obtain the asymptotic expansion of the thermal rate is the expansion of the partition function as an asymptotic series in $\hbar$,
\begin{align}
\label{eq:Zr_asymptotic}
Z_{\mathrm{r}} %
\sim Z_{\mathrm{r},0} \left[1+\hbar \GammaZr + \mathcal{O}(\hbar^2) \right].
\,
\end{align}
Here $Z_{\mathrm{r},0}$ is the leading-order approximation to the reactant partition function (e.g., translational or vibrational within the harmonic approximation), and $\GammaZr$ is the first-order correction (equivalent to VPT2\cite{Mills1972VPT2}). We explain how to evaluate these terms in Sec.~\ref{sec:Zr}.  

Combining the expansions from Eqs.~\eqref{eq:kZr_asymptotic} and \eqref{eq:Zr_asymptotic} yields the asymptotic series for the thermal rate up to first order 
\begin{equation}
\label{eq:k_asymptotic}
k %
\sim 
k_{\mathrm{inst}, 0}  \left[
1 + \hbar \Gamma_1 + \mathcal{O}(\hbar^2)
\right]
\,,
\end{equation}
where, $k_{\mathrm{inst}, 0}$, is the leading-order instanton rate 
(equivalent to the original theory of Miller),\cite{Miller1975semiclassical} which can be written in the present notation as
\begin{equation}
\label{eq:k_leading}
    k_{\mathrm{inst},0} %
    = \frac{1}{Z_{\mathrm{r},0}} \frac{A(\tilde{t})}{\hbar} \sqrt{\frac{2\pi\hbar}{S^{(2)}(\tilde{t})}}\,\e^{-S(\tilde{t})/\hbar}
    \,.
\end{equation}
Here $A(\tilde{t})$ and $S(\tilde{t})$ are the prefactor and action,  introduced in Eq.~\eqref{eq:cff_expansion}, evaluated at the stationary time, and are more precisely defined in Sec.~\ref{sec:cff}\@.
Furthermore, $S^{(2)}(\tilde{t})$ is the second derivative of $S(t)$ with respect to time at $t=\tilde{t}$ and is discussed in Sec.~\ref{sec:t_integration}.
Finally, the central focus of the present paper is the total first-order correction to the rate, $\Gamma_1$, which we see can be decomposed into three contributions
\begin{align}
\Gamma_1 = \Gammax + \Gammat - 
\GammaZr
\,.
\end{align}
the first two being the spatial correction and temporal correction from Eq.~\eqref{eq:kZr_asymptotic}, and the final term, $\GammaZr$, is the first-order correction from the reactant partition function, which contributes with a minus sign as $Z_{\mathrm r}$ appears in the denominator. %

Having obtained the asymptotic series for the rate up to first order, we can define the first-order corrected instanton rate as
\begin{equation}
k_{\mathrm{inst},1} =
k_{\mathrm{inst}, 0}  \left[
1 + \hbar \Gamma_1\right]
\,.
\end{equation}
This first-order approximation to the rate is asymptotic to the exact rate with error of $\mathcal{O}(\hbar^2)$. Additionally, the first-order correction itself is an error estimate for the leading-order rate. That is, if $\Gamma_1$ is small, we can assume that $k_{\mathrm{inst},0}$ is accurate. %
Moreover, using \emph{cumulant resummation}, we can obtain a potentially more accurate approximation $k_{\mathrm{inst}, 1\mathrm{c}}=k_{\mathrm{inst}, 0}  \exp\left[\hbar \Gamma_1\right]$; we explain this in more detail in Sec.~\ref{sec:cumulant_rate}.

Let us now summarize the procedure for finding the first-order corrected rate.
To evaluate both $k_{\mathrm{inst},0}$ and $\Gamma_1$ from Eq.~\eqref{eq:k_asymptotic}, we first need to locate the instanton trajectory. This has been discussed in great detail in previous work;\cite{InstReview} here we just state that the same trajectory is needed for the first-order corrections as for the leading-order result. Secondly, we need up to second derivatives of the PES along the instanton trajectory for the leading-order result and up to fourth derivatives for the first-order corrections.
With this, $k_{\mathrm{inst},0}$ and $\Gamma_1$ can be calculated using the formulas presented in Sec.~\ref{sec:cff}.

Note that although we derive the perturbative correction in two steps by first evaluating $\cff$ before integrating over time,
we could, in principle, evaluate both the spatial and temporal integrals simultaneously.
However, the uniqueness of asymptotic series guarantees that this would yield exactly the same result.

A true test of correctness which every asymptotic theory should aim to fulfill, is the asymptotic error test.
Namely, as $\hbar\to 0$, the relative error of instanton theory should follow a certain asymptotic trend.
By passing this test, we can be sure that we have a \emph{rigorously correct} perturbation theory in $\hbar$ of the formally exact quantum-mechanical expression, meaning that it tends to the exact result as $\hbar\to0$. %
We present the results of this test on an Eckart barrier in Sec.~\ref{sec:error_test}.

\section{Derivation of the theory}
\label{sec:cff}

In this section we give a more detailed exposition of the derivation,  providing explicit formulas for each term needed to compute the first-order rate as described in Sec.~\ref{sec:Overview_theory}. We begin by reviewing the derivation of the discretized path-integral expression for $\cff(t)$, defining notation necessary for later in the section. We then give the details of the asymptotic evaluation of this path integral that results in Eq.~\eqref{eq:cff_expansion}. Finally, we discuss the integral over time and review the complete information required to compute $\Gamma_1$. 

Following previous work,\cite{InstReview} we begin by expanding the trace in the position basis and inserting another resolution of the identity to give\cite{Miller1983rate}
\begin{align}
\label{eq:cffzero}
\cff({t}) &=
\frac{-\hbar^2}{4m^2}
\iint
L(x', x'', t)
\, 
\delta(\sigma(x'))
\, 
\delta(\sigma(x''))
\, \d x' \, \d x''
\,,
\end{align}
where 
\begin{align}
\label{eq:Axxt}
\!\!L(x', x'', t) \!&=\!
\left[\partial_{x'_\sigma}
K(x',x'',\ii t)\right]
\left[\partial_{x''_\sigma}
K(x'',x',\tot-\ii t)\right]
\nonumber\\&
-
\left[\partial_{x'_\sigma}
\partial_{x''_\sigma}
K(x',x'',\ii t)\right]
K(x'',x',\tot-\ii t)
\nonumber\\&
-
K(x',x'',\ii t)\left[
\partial_{x'_\sigma}
\partial_{x''_\sigma}
K(x'',x',\tot-\ii t)\right]
\nonumber\\&
+
\left[\partial_{x''_\sigma}
K(x',x'',\ii t)\right]
\left[\partial_{x'_\sigma}
K(x'',x',\tot-\ii t)\right]
.
\end{align}
Here $\partial_{x_\sigma} = \frac{\partial \sigma }{ \partial x} \cdot \, \partial_x$ 
is the directional derivative perpendicular to the dividing surface, and
\begin{align}
\label{eq:propagator}
K(x', x'', \ii t) &=
\bra{x'}
\e^{-i\hat{H}t/\hbar}
\ket{x''}
\,
\end{align}
are the \emph{propagators}.
Note that instanton theory only requires knowledge of the propagators at the stationary time, $t=\tilde{t}$, which is purely imaginary.
Hence, defining the propagator in this way and working in imaginary time $t=-\ii\tau$ is sufficient and we can write all our discretized observables in terms of $\tau$ without explicitly employing complex numbers.%

\subsection{Discretized path integrals}
\label{sec:discretized_PI}

In semiclassical theories, quantum-mechanical propagators [Eq.~\eqref{eq:propagator}] are usually approximated by van-Vleck propagators.\cite{vanVleck1928correspondence,GutzwillerBook}
Whilst this makes for a simple derivation of the leading-order theory,\cite{InstReview} the standard van-Vleck propagator does not contain information about first-order corrections. Instead, we prefer to make use of discretized path integrals,\cite{Feynman,Kleinert} which are more easily extensible and lead directly to the formulas used within ring-polymer instanton theory.

To obtain our path-integral expression for $\cff$, we begin by defining the two propagators appearing in Eq.~\eqref{eq:Axxt} as  
\begin{subequations}
    \begin{equation}
        K_a = K(x', x'', \tau_a) 
    \end{equation}
    \begin{equation}
        K_b = K(x'', x', \tau_b) \,,
    \end{equation}
\end{subequations}
with $\tau_a=\tau$ and $\tau_b=\tot -\tau$. Following the ring-polymer formalism, we can then express each propagator as an integral over a discretized path with $N_a$ or $N_b$ segments, such that, leaving the limit $N_a\to\infty$ and $N_b\to\infty$ implicit, we may write
\begin{subequations}
\label{eq:K_discretized}
    \begin{equation}
        K_{a} 
=
\left(
\frac{m}{2\pi \dta \hbar}
\right)^{N_af/2}
\int \d x_1\cdots\d x_{N_a-1}
\, \e^{-S_a(\vec{x}_a, \tau_a)/\hbar}
\,
    \end{equation}
    \begin{equation}
        K_{b} 
=
\left(
\frac{m}{2\pi \dtb \hbar}
\right)^{N_bf/2}
\int \d x_{N_a+1}\cdots\d x_{N-1}
\, \e^{-S_b(\vec{x}_b, \tau_b)/\hbar}
\,.
    \end{equation}
\end{subequations}
Here, $x_i$, are the ``beads'' of the ring polymer, which corresponds to snapshots of the system's position, and are separated by imaginary-time intervals of $\dta = \tau_a / N_a$ or $\dtb = \tau_b / N_b$.
The total number of beads is defined as $N=N_a+N_b$. We use the convention that  $x_0\equiv x_N\equiv x'$ and $x_{N_a}\equiv x''$ are the end beads, and $x_i$ for $i=1,\dots,N_a-1$ and $i=N_a+1,\dots,N-1$ are the intermediate beads, which are integrated over. 
The discretized action is given in each case by
\begin{subequations}
    \label{eq:two_fixed}
\begin{equation}
    \!\!\!\!S_{a}(\vec{x}_a, \tau_a)\! =\!
\sum_{i=0}^{N_a-1}\!\left[ \frac{m||x_{i+1}-x_i||^2}{2 \dta}
+
\dta
\frac{ V(x_i)+V(x_{i+1})}{2}
 \right]
\!\!\!\!
\end{equation}
\begin{equation}
   \!\!\!\! S_{b}(\vec{x}_b, \tau_b)\! =\!
\sum_{i=N_a}^{N-1}\!\left[ \frac{m||x_{i+1}-x_i||^2}{2 \dtb}
+
\dtb
\frac{ V(x_i)+V(x_{i+1})}{2}
 \right]
\,,
\end{equation}
\end{subequations}
where we denote the set of beads in each half-path as $\vec{x}_a=\{x_0,\dots,x_{N_a}\}$ and $\vec{x}_b=\{x_{N_a},\dots,x_N\}$; note that the beads $x'$ and $x''$ are included in both $\vec{x}_a$ and $\vec{x}_b$.
We can already see how discretization helps us: derivatives with respect to the end beads [cf.\ Eq.~\eqref{eq:Axxt}] are now trivial to evaluate by simply bringing the derivative inside the path integral of Eq.~\eqref{eq:K_discretized}.

Building on the discretized propagator we have just introduced, we can now combine the expressions for $K_a$ and $K_b$ together with the integrals over $x'$ and $x''$ in Eq.~\eqref{eq:cffzero} to give a single  discretized path-integral expression for $\cff(-\ii\tau)$.
To this end, we first define the full-trajectory action
\begin{align}
\label{eq:S_full}
S_N(\vec{x}, \tau) &= S_{a}(\vec{x}_a, \tau) + S_{b}(\vec{x}_b, \tot-\tau)
\,,
\end{align}
with the beads $\bm{x}=\{x_1,\dots,x_N\}$, which have cyclic indexing such that $x_0\equiv x_N$.
\newcommand{\preftwo}{\mathcal{N}}
With this we can write the final path-integral expression for $\cff(-\ii\tau)$ as
\begin{align}
\label{eq:cff_PI}
\cff (-\ii\tau) &=
\preftwo
\int \d \vec{x} \:
\delta(\sigma(x'))
\,
\delta(\sigma(x''))
\,
\Phi(\vec{x}, \tau)
\, \e^{-S_N(\vec{x}, \tau)/\hbar}
\,,
\end{align}
where 
\begin{equation}
\preftwo
= \frac{1}{4m^2}\left( \frac{m}{2\pi \hbar}\right)^{Nf/2}
\dta^{-N_af/2}
\,
\dtb^{-N_bf/2}
\end{equation}
and
\begin{align}
\label{eq:phi_term}
\notag
\Phi (\vec{x}, \tau) &=
(\partial_{x'_\sigma}
S_a)
(\partial_{x''_\sigma}
S_b)
-
(\partial_{x'_\sigma}
S_a)
(\partial_{x''_\sigma}
S_a)
\\&\quad -
(\partial_{x'_\sigma}
S_b)
(\partial_{x''_\sigma}
S_b)
+
(\partial_{x'_\sigma}
S_b)
(\partial_{x''_\sigma}
S_a)
\,.
\end{align}
Note, we will typically choose to divide the beads between the two half-trajectories such that $\dta=\dtb$ so that we can employ standard instanton-optimization techniques.

Finally, $\Phi(\vec{x},t)$ can be constructed using terms such as
\begin{align}
\label{eq:momentum_def}
-
\partial_{x'_\sigma}
S_a
&=
\frac{m(x_{1} - x_{0}) \cdot \vv{n} }{\dta}
\equiv
\pia(\vec{x}, \tau)
\,,
\end{align}
where $\vv{n}$ is a unit vector perpendicular to the dividing surface at $x_0$ (which, for simplicity, we assume to be a plane in Cartesian coordinates).
This term corresponds to the initial (imaginary-time) momentum perpendicular to the dividing surface on the $a$ half-trajectory.
Other terms in Eq.~\eqref{eq:phi_term} yield analogous momenta, which we describe in Appendix~\ref{app:momenta}. 
In this way, $\Phi(\vec{x}, \tau)$ can be expressed as
\begin{align}
\notag
\Phi (\vec{x}, \tau) &=
\pia(\vec{x}, \tau) \pib(\vec{x}, \tau)
+
\pia(\vec{x}, \tau) \pfa(\vec{x}, \tau)
\\
\quad
&+
\pib(\vec{x}, \tau) \pfb(\vec{x}, \tau)
+
\pfb(\vec{x}, \tau) \pfa(\vec{x}, \tau)
\,,
\end{align}
thus accounting for the momentum operators of Eq.~\eqref{eq:Flux_def}.

\subsection{Semiclassical approximation of the correlation function}
\label{sec:cff_inst}

In Eq.~\eqref{eq:cff_PI} we have introduced a path-integral expression for $\cff(-\ii\tau)$. 
Here we will use it to derive the first-order asymptotic expansion mentioned in Eq.~\eqref{eq:cff_expansion}, which will ultimately be used for computing the rate constant.

We make use of \emph{multivariate steepest descent}, a well-known technique from asymptotic analysis\cite{BenderBook,Wong1989Book} for integrals of the form
\begin{align}
\label{eq:I_def}
I(\hbar) &= \int g(\vec{x}) \, \e^{-S(\vec{x})/\hbar} \, \d \vec{x}
\,.
\end{align}
In this approach, one expands the integrand around the path $\tilde{\vec{x}}$ which minimizes $S$. Namely, $g(\vec{x})$ is expanded to the second order, $S(\vec{x})$ to the fourth order, and the exponentials of the third and fourth derivatives of $S(\vec{x})$ are likewise expanded.
This procedure then yields the following integral,
\begin{align}
\notag
I(\hbar)
&\sim
\int \d \vec{x} \:
\e^{-\left[S(\tilde{\vec{x}})+\frac{1}{2}\nabla_{ij}^2 S(\tilde{\vec{x}}) \Delta {x}^2_{ij}\right] / \hbar}
\\\notag&\phantom{\sim}\times
\left[
g(\tilde{\vec{x}})
+
\nabla_i g(\tilde{\vec{x}}) \Delta {x}_i
+
\frac{1}{2!}
\nabla^2_{ij} g(\tilde{\vec{x}}) \Delta {x}^2_{ij}
\right]
\\\notag
&\phantom{\sim}\times
\Big[
1
-
\frac{1}{3!\hbar}
\nabla^3_{ijk} S(\tilde{\vec{x}})
\Delta {x}^3_{ijk}
+
\frac{1}{2!(3!)^2\hbar^2}
\left(
\nabla^3_{ijk} S(\tilde{\vec{x}})
\Delta {x}^3_{ijk}
\right)^2
\\&\phantom{\sim}
\label{eq:I_def_expanded}
-
\frac{1}{4!\hbar}
\nabla^4_{ijkl} S(\tilde{\vec{x}})
\Delta {x}^4_{ijkl}
\Big]
\,,
\end{align}
where %
$\Delta{x}^n_{ij\dots} = (x_i-\tilde{x}_i)(x_j-\tilde{x}_j)\cdots$
and we make use of Einstein summation convention for repeated indices.
The result is then observed to be a sum of Gaussian integrals that can be evaluated using standard formulas.\cite{Kleinert}
For a more detailed explanation of multivariate steepest descent in the context of ring-polymer instanton theory, we refer the reader to Appendix B.2 of Ref.~\onlinecite{PCIT2023}.\footnote{In our previous work on tunneling splittings,\cite{PCIT2023} we have dealt with two cases. Thankfully, the present path integral corresponds to the simpler single-well integral and contains no zero- or imaginary modes which would make the evaluation more complicated.}
In addition to its use in this section, we also apply the steepest-descent procedure  to the reactant partition function in Sec.~\ref{sec:Zr} and (its one-dimensional version) to the time-integral in Eq.~\eqref{eq:Khbar}.

Applying multivariate steepest descent to the path integral in Eq.~\eqref{eq:cff_PI} we obtain Eq.~\eqref{eq:cff_expansion} along with explicit expressions for the prefactor $A(t)=A_N(\vec{\tilde{x}}, \tau)$ as
\begin{align}
\label{eq:Atau}
A_N(\vec{\tilde{x}}, \tau) &= \preftwo \hbar
\sqrt{\frac{(2\pi\hbar)^{Nf-2}}{\det \nabla^2 S_N(\tilde{\vec{x}}, \tau)}}
\, \Phi(\tilde{\vec{x}}, \tau)
\end{align}
and the action $S(t)=S_N(\tilde{\vec{x}}, \tau)$, i.e.,\ the ring-polymer action defined in Eq.~\eqref{eq:S_full} evaluated at the instanton trajectory $\tilde{\vec{x}}$. While the whole path integral spans over $N$ beads (and thus $Nf$ dimensions), two coordinates were pinned by the Dirac delta distributions arising from the flux operators. Therefore, the steepest-descent integration [and the accompanying application of $\nabla^2$ to $S_N(\tilde{\vec{x}}, \tau)$]  is conducted over the $Nf-2$ unpinned dimensions.

Noting that odd-powers of $\Delta x$ integrate to zero, the first-order correction to the correlation function can be expressed as a sum of a few terms,
\begin{equation}
    \label{eq:Gamma1x}
    \Gammax(t)= \Gamma^{(x)}_{\mathrm{A}} + \Gamma^{(x)}_{\mathrm{B}} + \Gamma^{(x)}_{\mathrm{C}}
\end{equation}
\addtocounter{equation}{-1}
with
\begin{subequations}
\begin{align}
\label{eq:Adef}
\!\!\Gamma^{(x)}_{\mathrm{A}} \!&=\! -\sum_{\mu\nu}\frac{3Q_{\mu\mu\nu\nu}}{4!D_{\mu\mu}D_{\nu\nu}}
\\
\label{eq:Bdef}
\!\!\Gamma^{(x)}_{\mathrm{B}} \!&=\!
\sum_{\mu\nu\rho}\frac{9T_{\mu\mu\nu}T_{\nu\rho\rho}}{2!(3!)^2D_{\mu\mu}D_{\nu\nu}D_{\rho\rho}}
+
\sum_{\mu\nu\rho}\frac{6T_{\mu\nu\rho}T_{\mu\nu\rho}}{2!(3!)^2D_{\mu\mu}D_{\nu\nu}D_{\rho\rho}}
\\
\label{eq:Gdef}
\!\!\Gamma^{(x)}_{\mathrm{C}} \!&=\! -\frac{3}{3!\Phi(\tilde{\vec{x}}, \tau)} \sum_{\mu\nu} \frac{\nabla_\mu \Phi T_{\mu\nu\nu}}{ D_{\mu\mu}D_{\nu\nu}}
+
\frac{1}{2\Phi} \sum_{\mu} \frac{\nabla_{\mu\mu}^2 \Phi }{D_{\mu\mu}}
\,,
\end{align}
\end{subequations}
where $D$, $T$ and $Q$ are the second, third and fourth derivative tensors of the action in normal-mode coordinates of $\nabla^2S_N$.
We also require spatial derivatives of $\Phi(\vec{x}, \tau)$ in normal-mode coordinates, which we provide in Appendix~\ref{app:momenta}.
All of these terms are evaluated at $\tilde{\vec{x}}$.

Numerical tests confirm that this first-order asymptotic expression correctly tends to the quantum-mechanical $\cff(-\ii\tau)$ quadratically as $\hbar\to0$. %
Note that four instantons are required to approximate the correlation function, formed of all combinations of left- and right-going trajectories.\cite{InstReview,QInst}
For brevity, we omit explicit illustration of these tests. Instead we confirm the correctness of our analysis by presenting the asymptotic dependence of the final rate expression in Sec.~\ref{sec:error_test}.

\subsection{Integration over time}
\label{sec:t_integration}

The next step in the procedure to obtain the rate is integrating the asymptotic expansion of $\cff(t)$ over $t$.
As we discussed in Sec.~\ref{sec:Overview_theory}, the initial integral [cf.\ Eq.~\eqref{eq:Khbar}] can be integrated using steepest descent to obtain the leading-order rate multiplied by a perturbation correction [Eq.~\eqref{eq:k_asymptotic}].
The one-dimensional version of the steepest-descent formula from Sec.~\ref{sec:cff_inst} can be used to obtain the contribution
\begin{equation}
\label{eq:Gammaxt}
    \Gammat = \Gamma^{(t)}_{\mathrm{A}} + \Gamma^{(t)}_{\mathrm{B}} + \Gamma^{(t)}_{\mathrm{C}}
\end{equation}\addtocounter{equation}{-1}
with
\begin{subequations}
\begin{align}
\Gamma^{(t)}_{\mathrm{A}} &= -\frac{3S^{(4)}(\tilde{t})}{4![S^{(2)}(\tilde{t})]^{2}}
\,,\\
\Gamma^{(t)}_{\mathrm{B}} &= \frac{15[S^{(3)}(\tilde{t})]^2}{2!(3!)^2[S^{(2)}(\tilde{t})]^{3}}
\,,\\
\Gamma^{(t)}_{\mathrm{C}} &= -\frac{A^{(1)}(\tilde{t})S^{(3)}(\tilde{t})}{2A(\tilde{t})[S^{(2)}(\tilde{t})]^{2}}
+
\frac{A^{(2)}(\tilde{t})}{2A(\tilde{t})S^{(2)}(\tilde{t})}
\,,
\end{align}
\end{subequations}
where, $S^{(n)}$ denotes the $n$-th derivative of $S$ with respect to real time (and similarly for $A^{(n)}$).

To obtain the derivatives of $A(t)$ and $S(t)$ with respect to $t$
we can use the Cauchy--Riemann condition\cite{ComplexVariables}
\begin{align}
    \label{eq:Cauchy_Riemann}
    \frac{\d}{\d t} = -\ii\frac{\d}{\d \tau} 
\end{align}
to express them in terms of more familiar imaginary-time derivatives of $S_N$. As a result, no real-time path-integral information is required to compute the instanton rate.
We thus require \emph{total} imaginary-time derivatives of the action up to fourth order. Using the chain rule (see Appendix~\ref{app:total_t}), these can be reformulated as combinations of various partial derivatives with respect to positions and imaginary time. Finally, the partial derivatives can be evaluated analytically starting from the original formula for the discretized action [Eqs.~\eqref{eq:two_fixed} and \eqref{eq:S_full}].

The additional first and second derivatives of $A(t)$ [Eq.~\eqref{eq:Atau}]
can then be obtained in a similar manner using the product rule, as outlined in Appendix~\ref{app:Taylor}.

We now have essentially all the information needed for obtaining the first-order corrected rate in Eq.~\eqref{eq:k_asymptotic}.
Ultimately all that is required beyond the standard ring-polymer instanton optimization\cite{InstReview} are third and fourth derivatives of the potential along the instanton pathway. These are inserted into the tensor contractions to obtain the correction factor $\Gamma_1$, and hence the perturbatively corrected instanton rate. 

The only remaining component we have not discussed is the first-order correction to the partition function $\GammaZr$. This can be straightforwardly evaluated following the same logic used above. The precise form of the correction depends on whether one is considering scattering (a bimolecular reaction) or escape from a metastable well (a unimolecular reaction). In the case of one-dimensional scattering the leading-order partition function is already exact, $Z_{\mathrm{r},0}=\Zr$. We, therefore, leave further discussion of the calculation of $\GammaZr$ to the multidimensional model considered in Sec.~\ref{sec:res_multidim}.   

\section{Results}\label{sec:results}

In this section, we apply the newly derived method to a number of systems of increasing complexity, starting with one-dimensional models before tackling multidimensional anharmonic problems.
In each case, we demonstrate that the perturbative correction systematically improves upon the standard leading-order instanton theory result.

\subsection{Symmetric Eckart barrier}
\label{sec:res_Eckart}

\newcommand{\EckParam}{\gamma}

As the first application of our method, we consider the symmetric Eckart barrier,\cite{Eckart}
\begin{align}
\label{eq:Eckart_sym}
V_{\mathrm{eck}}(x) &= 
\frac{V_0}{\cosh^2(x/a)}
\,.
\end{align}
For this simple problem, it is possible to calculate the exact quantum-mechanical rate as well as to derive an analytical formula for the first-order perturbative correction to the instanton approximation, as shown in Appendix~\ref{app:Eckart}.
Although we use these analytical formulas to calculate the rates in this section, we have confirmed that the numerical ring-polymer instanton approach gives identical results in the $N\to\infty$ limit; we demonstrate this in the supplementary material.

To facilitate comparison with previous results,\cite{Voth+Chandler+Miller, RPMDrefinedRate, RPInst, Faraday} we chose the parameters (in atomic units) $a=0.6604$, $m=1836$, $\EckParam = \pi\sqrt{2ma^2V_0} = 12$ (such that $V_0\approx0.00911$), and unless otherwise specified, $\hbar=1$.
The reciprocal crossover temperature for this barrier is $\betac =\EckParam / (V_0\hbar)$. %

\subsubsection{Rate dependence on temperature}

\begin{figure}[t]
    \centering
    \includegraphics{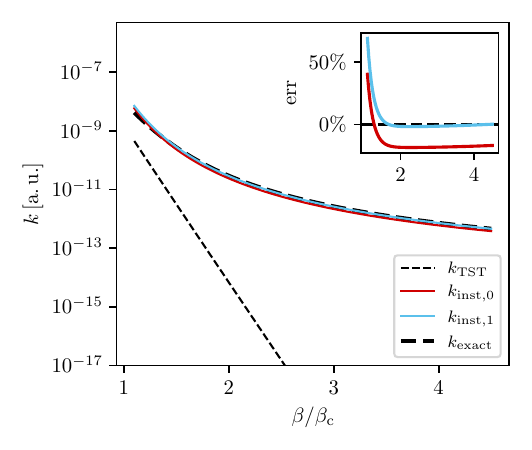}
    \caption{Arrhenius plot for a symmetric Eckart barrier with $\hbar=1$. Apart from an error in the vicinity of the crossover temperature ($\beta=\betac$), the first-order corrections offer a significant improvement over the lowest-order instanton method and closely match the exact rate in the deep-tunneling regime.}
    \label{fig:EckArr}
\end{figure}

We computed thermal reaction rates below the crossover temperature and present the resulting Arrhenius plot in Fig.~\ref{fig:EckArr}.
As expected, the TST rate cannot describe quantum tunneling and predicts a qualitatively wrong behavior.
By contrast, leading-order instanton theory captures the general trend as it includes tunneling effects and first-order corrected instanton theory captures more anharmonic effects and is thus in excellent agreement with the exact result.

More concretely, for $\beta>2\betac$ we see that the prediction of RPI+PC is consistently within $2\%$ of the exact rate on the range shown. This is significantly more accurate than the RPI result, which underestimates the rate by approximately $20\%$ over this range. However, as the crossover temperature is approached, $\beta \to_+ \betac$, we observe the error of both the leading-order and the first-order rate increase significantly. 
This is a well-understood limitation of instanton theory as one approaches the crossover temperature above which the barrier does not support a real instanton orbit.\cite{Affleck1981ImF,Benderskii,Faraday,JoeUniform} %

\subsubsection{The asymptotic error test}
\label{sec:error_test}

We have argued that our new theory is a rigorous $\hbar\to0$ asymptotic approximation to the true rate at a given value of $\tot=\beta\hbar$.
To demonstrate that this is the case,
we conduct a numerical test to determine the asymptotic behavior of the error.
Specifically, we calculate the instanton rates $k_{\mathrm{inst},0}(\tot, \hbar)$ and $k_{\mathrm{inst},1}(\tot, \hbar)$ as well as the exact rate $k$ for various values of $\hbar$.
Then we compute the relative errors %
\newcommand{\err}{\mathrm{err}}
\begin{equation}
\label{eq:err}
\err_n(\tot, \hbar) = \frac{k_{\mathrm{inst}, n}(\tot, \hbar)-k(\tot, \hbar)}{k(\tot, \hbar)}
\,.
\end{equation}
Considering Eq.~\eqref{eq:asymptotic_series} we see that, crucially, the uniqueness of asymptotic power series means that if and only if we have calculated all the terms up to order $n$ will we find that $\err_n(\tot, \hbar) \sim c \hbar^{n+1}$ for some constant $c$.

\begin{figure}[t]
    \centering
    \includegraphics{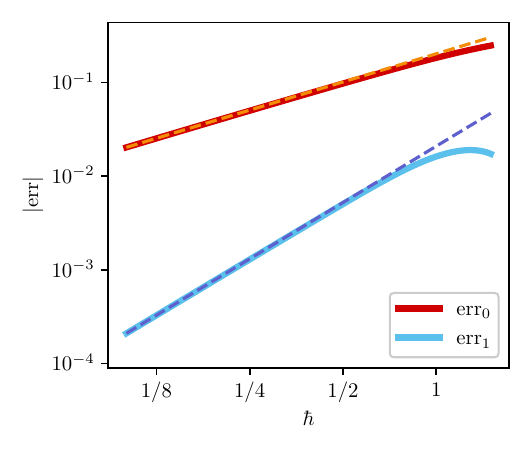}
    \caption{The relative error of the thermal rate at for symmetric Eckart barriers with varying $\hbar$ on a log--log plot. As $\hbar$ is varied, the total thermal time is kept constant at $\tot=2 \EckParam/V_0$.
    Dashed lines with a slope of $1$ and $2$, which represent the expected asymptotic behavior of $\err_0(\hbar)$ and $\err_1(\hbar)$, are shown.
    For small $\hbar$, we see that the leading-order error scales linearly with $\hbar$ and the first-order error scales quadratically, which confirms that the formulas have the correct asymptotic behavior. %
    Note that at $\hbar=1$, we recover the result from the Arrhenius plot in Fig.~\ref{fig:EckArr}. 
    }
    \label{fig:error_test}
\end{figure}

Varying $\hbar$ while keeping $\tot$ constant means that we select a reference inverse temperature $\beta_0$ and as we vary $\hbar$, we likewise vary the inverse temperature as $\beta(\hbar) = \beta_0/\hbar$.
Since the crossover temperature also depends on $\hbar$, the ratio $\beta(\hbar)/\betac(\hbar)$ stays constant throughout this process.
We present our asymptotic error test in Fig.~\ref{fig:error_test} for $\beta/\betac=2$ and indeed we observe that for $\hbar\to0$ the asymptotic behavior of both the leading-order and first-order instanton theories are correct.
The same behavior would be found for any fixed value of $\beta/\betac>1$.

Conducting the asymptotic error test is important, as it is a necessary condition to confirm that the theory is rigorously asymptotic as $\hbar\to0$ with $\tot$ kept constant.
Of course, for a given system it is possible that an alternative approximation scheme may be more accurate than the present theory.
However, the beauty of the present approach is that it can be applied to any system with the guarantee of asymptotic validity, i.e., it is a systematically improvable theory. This paper already offers a systematic improvement over the leading-order instanton theory and, although they would be costly to compute, other higher-order corrections could in principle be derived in a similar way within this framework.

\subsection{Asymmetric Eckart barrier}
\label{sec:res_asym}

Let us now consider a slightly more complicated system, the asymmetric Eckart barrier\cite{Eckart}
\begin{equation}
\label{eq:Eckart_asym}
V_{\text{as-eck}} (x) = \frac{V_0}{\cosh^2(x/a)} + 
\frac{V_\infty}{1+\exp(-2x/a)}
\,,
\end{equation}
where we use the standard parameters\cite{Pollak1998QTST,Jang1999QTST,RPInst,RPMDrefinedRate} (in reduced units $m=\hbar=k_{\mathrm{B}}=1$):
$V_\infty=-18/\pi$, $V_0=13.5/\pi$ and $a=8/\sqrt{3\pi}$, which give the reciprocal crossover temperature $\beta_\mathrm{c}=2\pi$.

\begin{figure}[t]
    \centering
    \includegraphics{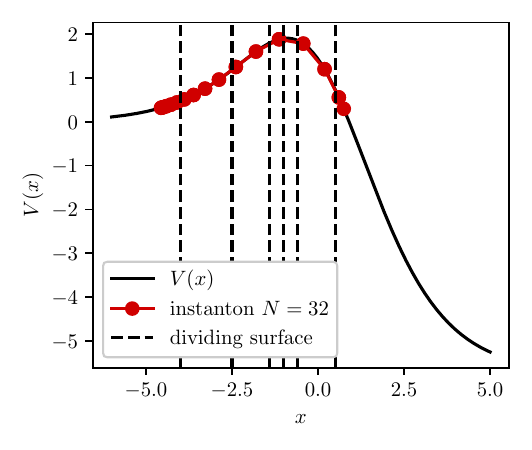}
    \caption{%
    A plot of the asymmetric Eckart PES [Eq.~\eqref{eq:Eckart_asym}] using standard parameters with several dividing surfaces $x^\ddagger$ and an instanton trajectory computed with $N=32$ beads.
    }
    \label{fig:EckAsym}
\end{figure}

For asymmetric systems, %
an important aspect to consider is the choice of dividing surface.
The exact rate is independent of the location of the dividing surface and leading-order instanton theory follows this principle correctly (as long as the dividing surface intersects with the instanton trajectory).\cite{AdiabaticGreens,InstReview,JoeUniform} %
In this section, we will demonstrate that the first-order correction also correctly satisfies dividing-surface independence.
The locations of the dividing surfaces $x^\ddagger$ that were considered are depicted in Fig.~\ref{fig:EckAsym}. In each case, we calculate the instanton rate at %
$\beta=2\betac$
and present the results in Table~\ref{table:EckAsym}.

\begin{table}[b]
\caption{%
Rate calculations at $\beta=2\betac$ for the asymmetric Eckart barrier [Eq.~\eqref{eq:Eckart_asym}] for various dividing surfaces, $x^\ddagger$.
The number of beads $N$ needed for convergence of $\hbar\Gamma_1$, individual components of the first-order corrections are displayed before being combined using 
$\hbar\Gamma_1 = \hbar\Gammax + \hbar\Gammat$. 
Although the individual components change significantly,
the sum of the first-order corrections remains independent of
$x^\ddagger$.
In the lower part of the table, the exact rate as well as the leading and first-order instanton rates are presented.
All calculations are shown in reduced units.
}
\label{table:EckAsym}
\begin{tabular}{d{2.1}d{4.0}  ccc}
\toprule
\head{$x^\ddagger$} & \head{$N$} & \head{$\hbar\Gammax/\%$} & \head{$\hbar\Gammat/\%$} & \head{$\hbar\Gamma_1/\%$} \\
\midrule
-4.0 & 4096 & $-20$   &  $30$     & 10 \\
-2.5 &  512 & $-22$   &  $32$     & 10 \\
-1.4 &   64 & $-45648$ & $45658$ & 10
\\
-1.0 &  128 & $1744$ & $-1734$   & 10 \\
-0.6 &  256 & $-8$   &   $18$    & 10
 \\
 0.5 & 4096 & $28$   & $-18$     & 10 \\
\midrule
\midrule
\head{} & \head{} & \head{$k_{\mathrm{inst}, 0}/10^{-8}$} & \head{$k_{\mathrm{inst}, 1}/10^{-8}$} & \head{$k_{\mathrm{exact}}/10^{-8}$}\\
\midrule
* & * & 3.674 & 4.04 & 4.056\\
\bottomrule
\end{tabular}
\end{table}

As expected, the leading- and first- order instanton rates are independent of the dividing surface; the leading-order theory roughly matches the exact rate with an error of $-9\%$ and the first-order theory is in excellent agreement with the exact rate with an error of $-0.4\%$.
Although we ultimately obtain the same result for each $x^\ddagger$, we note that the convergence to the $N\to\infty$ limit was observed to be slower with a poorly chosen dividing surface.

Moreover, we observed that for dividing surfaces $x^\ddagger\lesssim-1.2$, the nature of the right-going trajectories change.
In particular, they are no longer minima of the action, but saddle points (similarly to what was recently reported in a different context\cite{MeghnaCarbene,InstNearBarrier}).
However, the stationary-action principle still applies, as the action of a classical trajectory need not be a minimum.\cite{LandauMechanics}
The Hessian of the action will now have one extra negative eigenvalue,\footnote{This explains the high magnitude of the components $\hbar\Gammax$ and $\hbar\Gammat$.  As the point $x^\ddagger\approx-1.2$ is crossed, one eigenvalue becomes negative in a continuous way. Thus, near $x^\ddagger\approx-1.2$ this eigenvalue has a very low magnitude. It then appears in the denominator of Eq.~\eqref{eq:Gamma1x} in $D$, thus causing the high magnitude of the components.} but we simply take the absolute value of its determinant and use the negative eigenvalue for higher-order corrections.

At first sight, it appears unexpected that the components of the first-order correction $\Gammax$ and $\Gammat$ vary considerably with $x^\ddagger$. 
However, crucially, their sum $\Gamma_1$ is independent of $x^\ddagger$. Hence the final rate theory is independent of the dividing surface as it should be. 
This is all that is required of a physical rate theory: it highlights that the individual components do not have particular physical significance on their own and both the spatial and temporal contribution of the same order of $\hbar$ must be included to obtain a meaningful result.

\subsection{Multidimensional anharmonic model}
\label{sec:res_multidim}

So far, we have only studied one-dimensional systems where anharmonic effects occur along the instanton pathway. However, in many chemical reactions, the most important anharmonic effects are due to modes orthogonal to the reaction coordinate.
To illustrate the ability of our theory to treat multidimensional systems with anharmonicity perpendicular to the instanton pathway, we calculate the rate for the following two-dimensional system,
\begin{align}
\label{eq:PES2D}
V_{\mathrm{2D}}(x, y) &= V_{\mathrm{eck}}(x) + V_{\mathrm{M}}(y;\omega_{\mathrm{e}}, \chi_{\mathrm{e}}(x))
\,,
\end{align}
where $V_{\mathrm{eck}}(x)$ is the symmetric Eckart barrier [Eq.~\eqref{eq:Eckart_sym}] and $V_{\mathrm{M}}(y;\omega_{\mathrm{e}}, \chi_{\mathrm{e}})$ is the Morse potential
\begin{equation}
    V_{\mathrm{M}}(y;\omega_{\mathrm{e}}, \chi_{\mathrm{e}})= \frac{\omega_{\mathrm{e}}}{4\chi_{\mathrm{e}}} \left[1-\exp(- \sqrt{2 m \omega_{\mathrm{e}} \chi_{\mathrm{e}}}\,y)\right]^2.
\end{equation}
In order to couple the two degrees of freedom, we allow $\chi_{\mathrm{e}}(x)$ to vary along $x$,
\begin{align}
\label{eq:ChiGaussian}
\chi_{\mathrm{e}}(x) = \chi_{\mathrm{e},\infty} +
\left(\chi_{\mathrm{e},0} - \chi_{\mathrm{e},\infty} \right) \exp(-x^2/(2\sigma_{\mathrm{e}}^2))
\,,
\end{align}
such that the asymptotic value is $\chi_{\mathrm{e},\infty}$ and the value at $x=0$ is $\chi_{\mathrm{e},0}$.

\begin{figure}[t]
    \centering
    \includegraphics{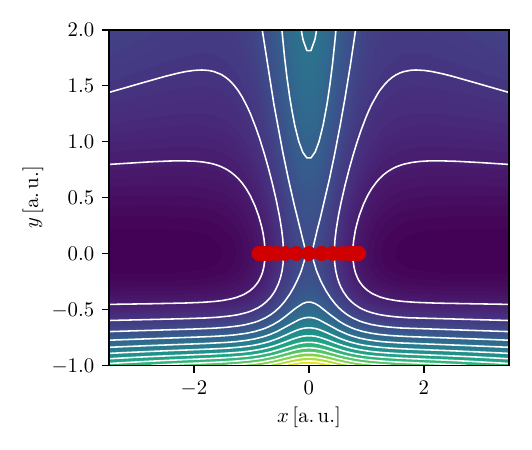}
    \caption{A contour plot of the coupled Eckart--Morse PES [Eq.~\eqref{eq:PES2D}] with %
    $\sigma_{\mathrm{e}}=4$.
    Contours are drawn at intervals of $0.003\,\mathrm{Ha}$.
    An instanton trajectory for $N=32$ beads at $\beta=2\betac$ is also included.
    }
    \label{fig:SpecPES}
\end{figure}

We %
use the same symmetric Eckart parameters as before (cf.\ Sec.~\ref{sec:res_Eckart}), and for the coupled Morse oscillator we use $\omega_{\mathrm{e}} = 0.0032$, $\chi_{\mathrm{e},\infty} = 0.01$, $\chi_{\mathrm{e},0} = 0.1$ with $\sigma_{\mathrm{e}}$ varied between $\sigma_{\mathrm{e}} = 1/4$ and $\sigma_{\mathrm{e}}=256$ (all in atomic units). %
We show the PES in Fig.~\ref{fig:SpecPES} for the intermediate case of $\sigma_{\mathrm{e}}=4$.
The instanton in this coupled system lies perfectly along $y=0$ and is identical to the instanton of the symmetric Eckart barrier along the $x$-axis.
The key difference is that the perpendicular modes now contribute anharmonic terms to the first-order perturbative correction.

The exact rates were calculated for comparison %
using standard quantum reactive scattering techniques. 
Specifically, the two-dimensional Born-Oppenheimer problem was transformed into an effective one-dimensional nonadiabatic problem in a crude adiabatic basis of Morse-oscillator eigenstates constructed at $x\to\infty$.
The resulting coupled-channel equations were then solved using the log-derivative method.\cite{Johnson1973logderivative,Mano1986logderivative}

\subsubsection{Reactant partition function}
\label{sec:Zr}

The coupled Eckart--Morse system is the first case presented in this paper with a nontrivial reactant partition function, $\Zr$. It can be expressed in the discretized path-integral notation of Sec.~\ref{sec:discretized_PI} as
\begin{align}
\nonumber
    Z_\mathrm{r} &= \Tr[\e^{-\tot\hat{H}/\hbar}\,\delta(\hat{x}-x_\mathrm{r})]
    \\ &= \preftwo
    \int \d \vec{x} \:
    \delta(x_{0}-x_\mathrm{r})
    \, \e^{-S_N(\vec{x}, \tau)/\hbar}
    \,,
\end{align}
where we take $x_\mathrm{r}\to-\infty$ in the reactant asymptote and  $S_N(\vec{x}, \tau)$ is the same as defined earlier [Eq.~\eqref{eq:S_full}] although here $\tau$ can be chosen arbitrarily, for example $\tau=\tot/2$.
To be consistent with the other approximations of our theory, we must compute the asymptotic approximation to this path integral using steepest descent [cf.\ Sec.~\ref{sec:cff_inst}]. 
The discretized action [Eq.~\eqref{eq:S_full}] can be used and the stationary path will be collapsed at the minimum of the Morse potential (with the $x$-coordinate of bead $0$ pinned at $x_\mathrm{r}$).
At leading order, one obtains the approximation
\begin{align}
    Z_{\mathrm{r},0} = \preftwo \sqrt{\frac{(2\pi\hbar)^{Nf-1}}{\det\nabla^2 S_N(\tilde{\vec{x}},\tau)}} 
    \,\e^{-S_N(\tilde{\vec{x}},\tau) /\hbar}
\end{align}
with the Hessian over $Nf-1$ unpinned beads.
Furthermore, we obtain the correction factor %
\begin{equation}
    \GammaZr = \Gamma^{(\mathrm{r})}_{\mathrm{A}} + \Gamma^{(\mathrm{r})}_{\mathrm{B}}\,,
\end{equation}
\addtocounter{equation}{-1}
with
\begin{subequations}
\begin{align}
\label{eq:Adef}
\!\!\Gamma^{(\mathrm{r})}_{\mathrm{A}} \!&=\! -\sum_{\mu\nu}\frac{3Q_{\mu\mu\nu\nu}}{4!D_{\mu\mu}D_{\nu\nu}}
\\
\label{eq:Bdef}
\!\!\Gamma^{(\mathrm{r})}_{\mathrm{B}} \!&=\!
\sum_{\mu\nu\rho}\frac{9T_{\mu\mu\nu}T_{\nu\rho\rho}}{2!(3!)^2D_{\mu\mu}D_{\nu\nu}D_{\rho\rho}}
+
\sum_{\mu\nu\rho}\frac{6T_{\mu\nu\rho}T_{\mu\nu\rho}}{2!(3!)^2D_{\mu\mu}D_{\nu\nu}D_{\rho\rho}}
\,.
\end{align}
\end{subequations}
Here, $D$, $T$ and $Q$ are the second, third and fourth derivative tensors of the action in normal-mode coordinates of $\nabla^2S_N$.

We note that in our case the $x$ and $y$ degrees of freedom decouple, such that the exact partition function is just a product of the translational partition function per unit length with the vibrational partition function of the Morse oscillator (with $\chi_{\mathrm{e}}=\chi_{\mathrm{e},\infty}$). %

We again confirmed the validity of the first-order correction using an asymptotic error test (equivalent to that used in Sec.~\ref{sec:error_test}). The exact partition function was computed using the known Morse oscillator eigenstates.

\subsubsection{First-order cumulant-corrected rate}
\label{sec:cumulant_rate}

So far we have only considered correcting the standard instanton rate using the simple partial sum  $k_{\rm inst,1}=k_{\rm inst,0}[1+\hbar \Gamma_1]$. However, it is well known that one can often obtain more accurate results using \emph{resummation} schemes.\cite{BenderBook} To illustrate this, consider the reactant partition function, for which a natural resummation scheme is obtained by considering the expansion of the free energy 
\begin{equation}
\begin{aligned}
    \label{eq:free-energy-def}
    F_{\rm r}(\tot;\hbar) = -\frac{\hbar} {\tot}\ln(\Lambda_{\rm th}\Zr)
    \end{aligned}
\end{equation}
rather than the partition function itself.
Note that we have multiplied $\Zr$ by the thermal de Broglie wavelength, $\Lambda_{\rm th}=\sqrt{2\pi\hbar\tot/m}$, to account for the known behavior of the translational contribution to the partition function. 

The free energy can then be expanded as a simple power series in $\hbar$. Matching term by term, one finds for this model that the coefficients of the power series are 
\begin{equation}
    F_{\rm r}(\tot;\hbar)\sim V_{\rm r} +\frac{\ln(2\sinh(\omega_{\rm e}\tot/2))}{\tot} \hbar - \frac{\GammaZr(\tot)}{\tot} \hbar^2 + \mathcal{O}(\hbar^3)\,,
\end{equation}
where $V_{\rm r}=0$ is the potential energy in the reactant minimum and the first-order term is recognized as the free energy of a harmonic oscillator.

Since even small changes in the free energy can lead to large changes in the partition function, the free-energy expansion typically leads to better results with a given number of terms.
Formally this corresponds to a \emph{cumulant resummation}, and is a standard technique from quantum field theory.\cite{PeskinSchroeder}
Truncating the free energy at first order in $\hbar$ becomes $Z_{\rm r, 0}$.
Truncating the free energy at $\hbar^2$ is equivalent to the first order cumulant resummation\footnote{With higher-order terms in the series for the free energy corresponding in the same manner to higher order cumulants of the original series for the correction to the partition function.}
\begin{equation}
    Z_{\rm r,1c}=Z_{\rm r,0}\exp[\hbar \GammaZr].
\end{equation}

Considering $\Gammax$ and $\Gammat$ as corrections to the
``transition-state'' or instanton partition function, it is natural to also expect a cumulant resummation of the rate,
\begin{equation}
\label{eq:k_cum}
k_{\mathrm{inst}, 1\mathrm{c}} =  k_{\mathrm{inst}, 0} \exp [\hbar \Gamma_1]
\,,
\end{equation}
 to typically give more accurate results than $k_{\rm inst,1}$.
For this reason, it is natural to favor cumulant resummation, rather than say Pad\'e resummation,\cite{BenderBook} as the most sensible choice of resummation scheme for reaction rates.

\subsubsection{Discussion of the results}

We present the results for the multidimensional anharmonic model in Table~\ref{table:SpecXeData}, with rates calculated at $\beta=2\betac$.
The first thing to highlight is that the leading-order instanton theory rates do not change as we vary $\sigma_\mathrm{e}$. This is because the leading-order theory does not account for anharmonicity perpendicular to the path and thus only depends on the frequency $\omega_\mathrm{e}$, which is constant in this model. In contrast, the first-order theory makes use of higher-order derivatives and can therefore account for the change in $\chi_{\rm e}$ along the path.
Consequently, we see that the first-order theory gives consistently more accurate predictions than the leading-order theory.
Specifically, we see that in going from the leading-order to the first-order instanton rate generally reduces the error from around 20--40\% to only 2--10\%. This is clearly a significant improvement. 

The first-order cumulant resummed approximation to the rate is even more accurate. We see that the errors are reduced to be as small as 0--3\%. The excellent behavior of the resummation in this case can be understood as arising from a combination of the accuracy of RPI+PC for the underlying one-dimensional Eckart barrier as seen in Fig.~\ref{fig:EckArr}, as well as the accuracy of the cumulant expansion for describing the partition function of a Morse oscillator. This reflects the fact that for a Morse oscillator the effects on the free energy from the higher derivatives are minimal. 

We note that the largest errors of 9.7\% for $k_{\rm inst,1}$ and 3.1\% for $k_{\rm inst,1c}$  occur at $\sigma_{\rm e}=4$. This corresponds to a local maximum of the exact quantum rate before it settles down to its plateau value in the limit of large $\sigma_{\rm e}$. In this plateau region the change in the anharmonicity is slow enough that the vibrational dynamics is adiabatic, and hence the transition-state theory picture is valid. The local maximum in the exact rate before this limit is caused by vibrational nonadiabatic effects. At this value of $\sigma_{\rm e}$, the anharmonicity changes significantly just beyond the end of the instanton. It is, therefore, unsurprising that instanton theory is unable to capture the subtle dynamical effect of the stiffening of the potential in this region and instead it simply predicts a monotonic rise in the rate.  
For smaller values of $\sigma_{\rm e}$, the dynamics are also vibrationally nonadiabatic, but because the anharmonicity changes within the region sampled by the instanton, it is correctly captured by the theory.

Finally, we highlight a point that may at first seem trivial, but is in fact central to the power of the present approach. The magnitude of $\hbar\Gamma_1$ is a rigorous error estimate for the leading-order theory. This is a feature of the fact that the theory is systematically improvable and $\hbar\Gamma_1$ is a perturbative correction. 
Drawing a connection with electronic-structure theory, it therefore closely resembles the difference say between CCSD and CCSD(T), or between Hartree--Fock and MP2. 
This is in contrast to theories which may often be more accurate, but are not systematically improvable, such as hybrid functionals with empirical parameters in density functional theory.

\begin{table}[t]
\caption{Rate calculations for the coupled Eckart--Morse PES [cf.\ Eq.~\eqref{eq:PES2D}] for various values of $\sigma_{\mathrm{e}}$, which controls the anharmonicity of the system.
Furthermore, the values for the uncoupled one-dimensional Eckart potential are included for reference (in the first row).
All rate constants are in atomic units ($\times 10^{-11}$) and were calculated at $\beta=2\betac$ with $N=1024$ beads.
While the leading-order theory cannot account for anharmonicity and thus has the same prediction
$k_{\mathrm{inst}, 0} =2.40\times 10^{-11}\,\mathrm{a.\,u.}$
for every value of $\sigma_{\mathrm{e}}$, the first-order theory and its cumulant resummation capture the correct behavior.
}
\label{table:SpecXeData}
\begin{tabular}{c d{1.2}d{1.2}d{1.2} d{2.0}d{2.0}d{1.1}d{1.1}}
\toprule
\head{$\sigma_{\mathrm{e}}$} &
\head{$k_{\mathrm{inst}, 1}$} &
\head{$k_{\mathrm{inst}, 1\mathrm{c}}$} &
\head{$k_{\mathrm{exact}}$}  &
\head{$\hbar\Gamma_1 / \%$}  & 
\head{$\err_0 / \%$}  & 
\head{$\err_1 / \%$}  &
\head{$\err_{1\mathrm{c}} / \%$}
\\
\midrule
--     & 2.88 & 2.95 & 2.94 & 20 & -18 & -2.0 & 0.2 \\
$1/4$  & 2.92 & 3.00 & 3.00 & 22 & -20 & -2.6 & 0.0 \\
$1/2$  & 3.08 & 3.21 & 3.18 & 28 & -25 & -3.1 & 1.0 \\
1      & 3.24 & 3.44 & 3.42 & 35 & -30 & -5.3 & 0.4 \\
2      & 3.30 & 3.53 & 3.52 & 38 & -32 & -6.2 & 0.2 \\
4      & 3.32 & 3.56 & 3.68 & 38 & -35 & -9.7 & -3.1 \\
8      & 3.33 & 3.56 & 3.57 & 39 & -33 & -6.8 & -0.2 \\
16     & 3.35 & 3.56 & 3.61 & 40 & -34 & -7.3 & -1.4 \\
32     & 3.35 & 3.56 & 3.63 & 40 & -34 & -7.7 & -1.8 \\
64     & 3.35 & 3.56 & 3.63 & 40 & -34 & -7.9 & -2.0 \\
128    & 3.35 & 3.56 & 3.64 & 40 & -34 & -7.9 & -2.0 \\
256    & 3.35 & 3.56 & 3.64 & 40 & -34 & -7.9 & -2.0 \\
\bottomrule
\end{tabular}
\end{table}

\subsection{Collinear \HHtwo}

The final system we investigate is the collinear \HHtwo reaction and its isotopic variants \DDtwo and \TTtwo. In these systems, the nuclear masses are scaled by a constant factor, allowing us to analyze the impact of this scaling on the accuracy of the RPI+PC method.
We use the complete configuration interaction (CCI) potential energy surface of Ref.~\citenum{ccipot} in Jacobi coordinates.
Exact results were calculated using a sinc-DVR\cite{Colbert1992DVR} (with a potential energy cutoff) by numerical integration of the flux--flux correlation function up to the plateau time, with the dividing surface chosen to pass through the saddle-point of the potential.

Let us first explain our coordinate system.
Given the coordinates $\mathbf{r}_n$ of the individual atoms, we can define the 
\emph{Jacobi coordinates}
$\mathbf{R} = \mathbf{r}_3 - (\mathbf{r}_1-\mathbf{r}_2)/2$ and $\mathbf{r} = \mathbf{r}_1-\mathbf{r}_2$ together with their associated masses $m_R=2m_{\mathrm{X}}/3$ and $m_r=m_{\mathrm{X}}/2$, where $m_{\mathrm{X}}$ is the mass of the hydrogen isotope $\mathrm{X}\in \{\mathrm{H}, \mathrm{D}, \mathrm{T}\}$.
The magnitudes of the Jacobi coordinates $R$ and $r$ are then sufficient to define the system's configuration under the restriction that all atoms are on a line.

\begin{figure}[t]
    \centering
    \includegraphics{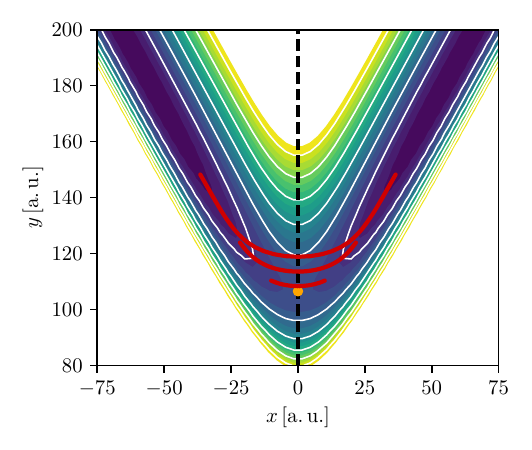}
    \caption{A contour plot of the mass-weighted $\HHtwo$ PES in atomic units with instanton trajectories (red) at $100\,\mathrm{K}$, $200\,\mathrm{K}$ and $300\,\mathrm{K}$.
    Contours are drawn at intervals of $0.0112\,\mathrm{Ha}$.
    The PES is symmetric around the the dividing surface (dashed black line).
    As the temperature increases, the total thermal time $\tot$ decreases and the instanton trajectories become shorter, moving closer to the transition state, marked by an orange point.%
    }
    \label{fig:PES_H2}
\end{figure}

Next,  we transform into mass-weighted coordinates $R \sqrt{m/m_R}$ and $r\sqrt{m/m_r}$, where $m=1$ a.\,u.\ is the reference mass. Lastly, we transform into our final coordinates $(x, y)$ by rotating our coordinate system by $\pi/3$ such that the dividing surface $x=0$ becomes the line of reflection of our system.
The resulting PES for the $\HHtwo$ reaction is shown in Fig.~\ref{fig:PES_H2}.

While the original PES is the same for all three isotopic variants, the mass-weighted coordinate transformation changes the length scale of each PES differently. As a result, the surfaces of all three reactions have the same geometric shape, but differ by a constant scaling factor $\sqrt{m_{\mathrm{H}} / m_\mathrm{X}}$.
This leads to different crossover temperatures, $\Tc^{(\HHtwo)} = 343.86\,\mathrm{K}$, $\Tc^{(\DDtwo)}=243.24\,\mathrm{K}$ and $\Tc^{(\TTtwo)}=198.77\,\mathrm{K}$.

\begin{figure*}[t]
    \centering
    \includegraphics{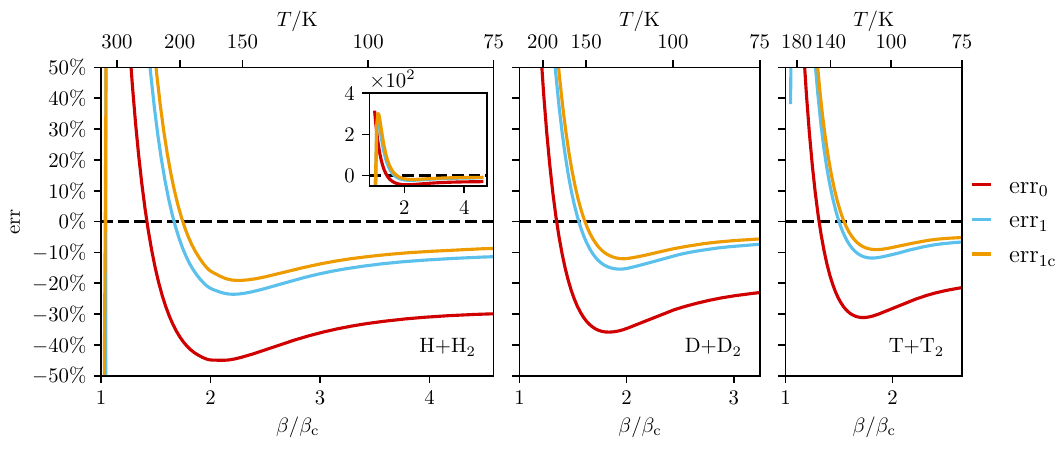}
    \caption{
    The relative errors [Eq.~\eqref{eq:err}] of instanton rates depending on temperature for $\HHtwo$, $\DDtwo$ and $\TTtwo$.
    The \HHtwo plot additionally contains an inset showing the full range of errors. %
    Apart from the region near the crossover temperature, the first-order rate is more accurate than the leading-order rate, and the cumulant-corrected rate represents a further improvement.
    Moreover, the error of the instanton rate decreases as the isotopic mass increases.
    }
    \label{fig:HDT_plot}
\end{figure*}

We computed instanton rates for all three systems across a range of temperatures.
We show the relative errors of the instanton calculations compared to the DVR results in Fig.~\ref{fig:HDT_plot}.
All three systems exhibit behavior similar to that observed for the symmetric Eckart barrier [Sec.~\ref{sec:res_Eckart}].  With the exception of the region near the crossover temperature, the first-order corrected rates consistently improve upon the leading-order results, with the cumulant-corrected rates performing even better. In all cases, the relative errors are seen to begin to plateau once firmly in the deep-tunneling regime, i.e., when $\beta>2\betac$.

In the region near the crossover temperature, neither the leading-order nor the first-order correction predicts the rate accurately. We show this behavior fully in the inset of the \HHtwo reaction in Fig.~\ref{fig:HDT_plot}. This breakdown of instanton theory is well-known\cite{Affleck1981ImF,Haenggi1988crossover,Cao1996QTST,Benderskii,Faraday,Kryvohuz2011rate}
and in future work, we aim to remedy this issue with the uniform instanton theory approach of Ref.~\citenum{JoeUniform}.

Comparing across isotopic variants, the accuracy of the rates systematically improves as mass increases.
This can be understood by noting that mass and $\hbar$ appear together in the Schr\"odinger equation as $\hbar^2/m_{\mathrm{X}}$, such that 
an error of order $\mathcal{O}(\hbar)$ is equivalent to one of order $\mathcal{O}(1/\sqrt{m_\mathrm{X}})$.
Consequently, the systematic improvement can be attributed to the better performance of instanton approaches in the semiclassical limit.
Cumulant resummation offers a further improvement over the standard partial sum here.  However the improvement is not as substantial as in the coupled Eckart--Morse case. This reflects the fact that the primary effect here is the change to the frequency, not the anharmonicity perpendicular to the path. 

To investigate the trends, we compare relative errors at a fixed ratio\footnote{Varying the mass fundamentally changes the system we are studying, such that we need to hold constant the ratio $\beta/\betac$ (instead of $\tot$) to observe asymptotic behavior.} of $\beta/ \betac=2$ in Table~\ref{table:HpH2_half}.
The relative error decreases as mass increases, and according to the analysis in Sec.~\ref{sec:error_test}, we expect the first-order corrected rate to improve faster than the leading-order rate.
Indeed, while the error of the leading-order rate decreases roughly linearly with $1/\sqrt{m_{\mathrm{X}}}$, the error of the first-order rate decreases more rapidly (although for these values it is not yet well fit by the known limiting quadratic decay; this would require going to larger masses).
Finally, although our theory does not predict how the error of the cumulant-corrected rate should decay, the error still improves over the first-order rate and also decreases superlinearly as $1/\sqrt{m_{\mathrm{X}}}\to0$.
\begin{table}[b]
\caption{%
Relative errors of instanton rates at $\beta = 2\betac$ for the \HHtwo, \DDtwo and \TTtwo reactions.
}
\label{table:HpH2_half}
\begin{tabular}{c d{1.2}d{3.0}d{3.0}c}
\toprule
\head{Reaction} & \head{$\sqrt{m_{\mathrm{H}}/m_{\mathrm{X}}}$} & \head{$\err_0/\%$} & \head{$\err_1/\%$} & \head{$\err_{1\mathrm{c}}/\%$} \\
\midrule
\HHtwo & 1.00 & -45 &  -22 & $-16$ \\
\DDtwo & 0.71 & -35 &  -15 & $-12$ \\
\TTtwo & 0.58 & -28 &  -11 & $-8.4$ \\
\bottomrule
\end{tabular}
\end{table}

\section{Conclusion}
\label{sec:conclusion}

In this work, we have introduced the first-order perturbative correction to instanton rate theory, which we have derived in the ring-polymer framework (RPI+PC)\@.
The new RPI+PC theory combines the advantages of SCTST with semiclassical instanton theory to account for anharmonicity in addition to deep tunneling, which makes it a uniquely accurate, yet affordable, theory. As the only information needed to obtain the RPI+PC rate is the ordinary instanton trajectory and local derivatives along that trajectory, this theory promises to be applicable to larger molecular systems and compatible with on-the-fly electronic-structure calculations using the framework constructed for SCTST and VPT2 calculations.\cite{Nguyen2010SCTST}
Alternatively, modern machine-learning approaches to fitting potential energy surfaces offer high-order derivatives at a low cost.\cite{TransferLearning,TransferLearning2}

We have confirmed that RPI+PC truly is a rigorous asymptotic expansion of the exact quantum-mechanical rate as $\hbar\to0$ (with $\tot$ kept constant) using formal tests of the asymptotic error. %
As a result, our study of a range of model systems including the collinear \HHtwo, \DDtwo and \TTtwo reactions showed a systematic improvement
over the leading-order RPI rates.%
In addition to giving a more accurate result, the perturbative correction can also be used to estimate the error made by instanton theory due to neglect of anharmonicity.
As a consequence of the systematic and rigorous nature of the theory, a small value of $\Gamma_1$ indicates that the standard instanton rate is already very accurate. On the other hand, a large value of $\Gamma_1$ suggests that anharmonicity is important.
This behavior was illustrated clearly in our two-dimensional Eckart--Morse model.

In future work, surface chemistry could already be tackled with the present theory, similarly to standard instanton theory,\cite{Jonsson2011surface,Hgraphene} but more work is needed to describe full-dimensional gas-phase reactions.
In this case, rotation symmetry introduces zero-modes into the ring-polymer Hessian, which need to be properly accounted for. 
We have already treated a zero mode in the context of tunneling splitting\cite{PCIT2023} and assume that the treatment of rotational zero-modes in the present case will be possible in a similar way.

Furthermore, the rigor of our perturbation theory enables RPI+PC to be combined with other systematic improvements to instanton theory.
An important next extension will be to combine RPI+PC with uniform instanton theory to remedy the crossover-temperature problem.\cite{JoeUniform}
Additionally, 
ring-polymer instanton theory can be generalized to treat nonadiabatic reactions,\cite{PhilTransA,GRperspective} even in the inverted regime.\cite{inverted}
A similar derivation can be followed to obtain a perturbative correction for golden-rule instanton theory, providing the same advantages as in the present adiabatic case.

\section*{Acknowledgments}
JEL was supported by an Independent Postdoctoral Fellowship at the Simons Center for Computational Physical Chemistry, under a grant from the Simons Foundation (839534, MT).
The authors acknowledge financial support from the Swiss National Science Foundation through SNSF project 207772. 
We thank Curdin Keller for testing some ideas during the early investigatory stages of this project.

\section*{Data availability}

The data that support the findings of this study are available within the article and its supplementary material.

\appendix

\section{Total time derivatives of the action}
\label{app:total_t}

In this Appendix, we present the formulas for the total time derivatives of $S(\vec{x}(t), t)$ using Einstein summation convention:
\begin{align}
\frac{\d S}{\d t} &= 
\frac{\partial S}{\partial t}
+
\frac{\partial S}{\partial x_i}
\frac{\d x_i}{\d t}
\\
\frac{\d^2 S}{\d t^2} &=
\frac{\partial^2 S}{\partial t^2}
+
2\frac{\partial^2 S}{\partial t \partial x_i} \frac{\d x_i}{\d t}
+
\frac{\partial^2 S}{\partial x_i\partial x_j}
\frac{\d x_i}{\d t}
\frac{\d x_j}{\d t}
+
\frac{\partial S}{\partial x_i} \frac{\d^2 x_i}{\d t^2}
\\
\label{eq:d3Sdt3}
\nonumber
\frac{\d^3 S}{\d t^3} &=
\nonumber
\frac{\partial^3 S}{\partial t^3}
+
3
\frac{\partial^3 S}{\partial t^2 \partial x_i} \frac{\d x_i}{\d t}
+
3\frac{\partial^3 S}{\partial t \partial x_i\partial x_j}
\frac{\d x_i}{\d t}
\frac{\d x_j}{\d t}
\\&\phantom{=}
\nonumber
+
3\frac{\partial^2 S}{\partial t \partial x_i} \frac{\d^2 x_i}{\d t^2}
+
\frac{\partial^3 S}{\partial x_i\partial x_j\partial x_k}
\frac{\d x_i}{\d t}
\frac{\d x_j}{\d t}
\frac{\d x_k}{\d t}
\\&\phantom{=}
+
3
\frac{\partial^2 S}{\partial x_i\partial x_j}
\frac{\d^2 x_i}{\d t^2}
\frac{\d x_j}{\d t}
+
\frac{\partial S}{\partial x_i} \frac{\d^3 x_i}{\d t^3}
\\
\nonumber
\frac{\d^4 S}{\d t^4} &= 
\frac{\partial^4 S}{\partial t^4}
+
4
\frac{\partial^4 S}{\partial t^3 \partial x_i} \frac{\d x_i}{\d t}
+
6
\frac{\partial^4 S}{\partial t^2 \partial x_i\partial x_j}
\frac{\d x_i}{\d t}
\frac{\d x_j}{\d t}
\\&\phantom{=}
\nonumber
+
6
\frac{\partial^3 S}{\partial t^2 \partial x_i}
\frac{\d^2 x_i}{\d t^2}
+
4\frac{\partial^4 S}{\partial t \partial x_i\partial x_j\partial x_k}
\frac{\d x_i}{\d t}
\frac{\d x_j}{\d t}
\frac{\d x_k}{\d t}
\\&\phantom{=}
\nonumber
+
12\frac{\partial^3 S}{\partial t \partial x_i\partial x_j}
\frac{\d^2 x_i}{\d t^2}
\frac{\d x_j}{\d t}
+
4\frac{\partial^2 S}{\partial t \partial x_i} \frac{\d^3 x_i}{\d t^3}
\\&\phantom{=}
\nonumber
+
\frac{\partial^4 S}{\partial x_i\partial x_j\partial x_k\partial x_l}
\frac{\d x_i}{\d t}
\frac{\d x_j}{\d t}
\frac{\d x_k}{\d t}
\frac{\d x_l}{\d t}
\\&\phantom{=}
\nonumber
+
6
\frac{\partial^3 S}{\partial x_i\partial x_j\partial x_k}
\frac{\d^2 x_i}{\d t^2}
\frac{\d x_j}{\d t}
\frac{\d x_k}{\d t}
+
3
\frac{\partial^2 S}{\partial x_i\partial x_j}
\frac{\d^2 x_i}{\d t^2}
\frac{\d^2 x_j}{\d t^2}
\\&\phantom{=}
+
4
\frac{\partial^2 S}{\partial x_i\partial x_j}
\frac{\d^3 x_i}{\d t^3}
\frac{\d x_j}{\d t}
+
\frac{\partial S}{\partial x_i}
\frac{\d^4 x_i}{\d t^4}
\,,
\end{align}
where the partial derivatives of $S(\vec{x}, t)$ can be evaluated analytically using Eqs.~\eqref{eq:two_fixed}, \eqref{eq:S_full} and the Cauchy--Riemann condition [Eq.~\eqref{eq:Cauchy_Riemann}].

Additionally, we need the total time derivative of $\vec{x}(t)$.
In our case, we will always evaluate the action at the stationary point $\vec{\tilde{x}}$, though we need to know how it changes when we change $\tau$.
We could in principle obtain derivatives of $\vec{\tilde{x}}(\tau)$ numerically by re-optimising the trajectory at different values of $\tau$.
However, a more efficient approach is to use analytical formulas, which can be derived by implicit differentiation of the equation defining $\vec{\tilde{x}}$. This approach is also used in Ref.~\onlinecite{Kleinert} and Ref.~\onlinecite{GoldenRPI}.

For example, the first derivative can be obtained as follows.
First, we take the gradient of the action defined in Eq.~\eqref{eq:S_full} and set it to zero
\begin{align}
\label{eq:GradS}
\nabla
S(\vec{\tilde{x}}, \tau)  &= 0
\end{align}
This gives an equation with $\vec{\tilde{x}}$ as its solution.
Then, we differentiate Eq.~\eqref{eq:GradS} with respect to $\tau$ using the chain rule to obtain the linear equation
\begin{align}
\frac{\partial^2 S(\vec{x}, \tau)}{\partial \vec{x}^2}\Big\lvert_{\vec{x}=\tilde{\vec{x}}}
\frac{\d \tilde{\vec{x}}}{\d \tau}
+
\frac{\partial^2 S(\vec{x}, \tau)}{\partial \tau \partial \vec{x}}\Big\lvert_{\vec{x}=\tilde{\vec{x}}}
&= 0
\,.
\end{align}
This can then be solved for $\d \vec{\tilde{x}} / \d \tau$.
For higher derivatives, one can repeat the procedure.

\section{Derivatives of Prefactors}
\label{app:Taylor}

To calculate the first-order corrected rate, we need the spatial derivatives of $\Phi(\vec{x}, \tau)$ [Eq.~\eqref{eq:phi_term}] and the time derivatives of $A_N(\tau)$ [Eq.~\eqref{eq:Atau}].
The required derivatives of $\Phi(\vec{x}, \tau)$ are derived in Appendix~\ref{app:momenta}, while those of $A_N(\tau)$ are obtained in Appendix~\ref{app:determinant}.

\subsection{Momenta}
\label{app:momenta}

The flux operator gives rise to the $\Phi(\vec{x}, \tau)$ term, which can be related to the sum of products of momenta [Eq.~\eqref{eq:momentum_def}].
In this section, we will explain how to calculate derivatives of the momenta, assuming that the dividing surface is a plane in Cartesian coordinates. Subsequently, the derivatives of $\Phi(\vec{x}, \tau)$, which are needed for Eq.~\eqref{eq:Gdef} and in Appendix~\ref{app:determinant}, can be obtained in a straightforward way.

The initial and final momenta on both half-trajectories are are needed
\begin{subequations}
\begin{align}
\label{eq:pia}
\pia(\vec{x}, \tau)
&\equiv
\frac{m(x_{1} - x_{0})\cdot\vv{n}}{\dta}
\\
\pfa(\vec{x}, \tau)
&\equiv
\frac{m(x_{N_a} - x_{N_a-1})\cdot\vv{n}}{\dta}
\\
\pib(\vec{x}, \tau)
&\equiv
\frac{m(x_{N_a+1} - x_{N_a})\cdot\vv{n}}{\dtb}
\\
\pfb(\vec{x}, \tau)
&\equiv
\frac{m(x_{N}-x_{N-1})\cdot\vv{n}}{\dtb}
\,,
\end{align}
\end{subequations}
where $\vv{n}$ is a unit vector perpendicular to the dividing surface at $x_0$.
To avoid repetition, we will only present the derivatives of $\pia(\vec{x},\tau)$ here.

Let us start by writing the general formula for the total time derivatives of the momentum
$\pia(\vec{\tilde{x}}(\tau), \tau)$%
\begin{align}
\frac{\d \pia}{\d t} 
&=
\frac{\partial \pia}{\partial t} 
+
\frac{\partial \pia}{\partial \vec{x}}
\cdot
\frac{\d \vec{x}}{\d t}
\\
\frac{\d^2 \pia}{\d t^2} 
&=
\frac{\partial^2 \pia}{\partial t^2} 
+
2\frac{\partial^2 \pia}{\partial t \partial \vec{x}} 
\!\cdot\!
\frac{\d \vec{x}}{\d t}
+
\frac{\d \vec{x}}{\d t}
\!\cdot\!
\frac{\partial^2 \pia}{\partial^2 \vec{x}^2}
\!\cdot\!
\frac{\d \vec{x}}{\d t}
+
\frac{\partial \pia}{\partial \vec{x}}
\!\cdot\!
\frac{\d^2 \vec{x}}{\d t^2}
\,,
\end{align}
where the total derivatives of $\vec{x}$ at the point $\vec{\tilde{x}}(\tau)$ were described in Appendix~\ref{app:total_t}.

Next, we need to evaluate the individual derivatives using the definition of the momentum provided in Eq.~\eqref{eq:pia} and the Cauchy--Riemann condition [Eq.~\eqref{eq:Cauchy_Riemann}].
First, the partial time derivative is
\begin{align}
\frac{\partial}{\partial t} \pia 
&=
-\frac{-\ii}{\tau} \pia  
\,,&
\frac{\partial^2}{\partial t^2} \pia 
&=
\frac{2(-\ii)^2}{\tau^{2}} \pia  
\,.
\end{align}
The partial position derivative is
\begin{align}
\label{eq:p_spatial}
\frac{\partial}{\partial {x}_j} \pia
&=
\frac{m}{\dta} \delta_{1,j}
\,,&
\frac{\partial^2 }{\partial {x}_j^2} \pia  &= 0  \,,
\end{align}
where $\delta_{1,j}$ is the Kronecker delta.
Lastly, the mixed derivative is
\begin{align}
\frac{\partial}{\partial t \partial {x}_j} \pia  &= \ii\frac{m}{\tau\dta} \delta_{1,j} \,.
\end{align}

With the above equations, we can evaluate the total time derivatives of $\Phi(\vec{x}, \tau)$ needed for the total time derivative of $A_N(\tau)$ in Appendix~\ref{app:determinant}.
Furthermore, Eq.~\eqref{eq:p_spatial} 
combined with Eqs.~\eqref{eq:phi_term} and Eq.~\eqref{eq:momentum_def} defines the spatial derivatives of $\Phi(\vec{x}, \tau)$. We can transform these derivatives into normal-mode coordinates, which is required for the evaluation of Eq.~\eqref{eq:Gdef}.

\subsection{Determinant}
\label{app:determinant}

\newcommand{\ApartI}{\Omega}
\newcommand{\ApartII}{J}

We can rewrite $A_N(\tau)$ [Eq.~\eqref{eq:Atau}] as follows:
\begin{align}
A_N(\tau) &=
c\times
\ApartI(\tau)
\times
\Phi(\tau)
\end{align}
with
\begin{subequations}
\begin{align}
c &= \frac{(2\pi)^{(Nf-2)/2}}{4m^2}\left( \frac{m}{2\pi }\right)^{Nf/2}
\\
\ApartI(\tau) &= \sqrt{\frac{1}{\dta^{N_af} \dtb^{N_bf}\det \ApartII}}
\\
\ApartII &=  \nabla^2 S_N(\tau)
\,.
\end{align}
\end{subequations}
Then we can express real-time derivatives of $A_N(\tau)$ using the product rule:
\begin{subequations}
\label{eq:An}
\begin{align}
A^{(1)}({\tau}) &= c 
\left(
\ApartI^{(1)}(\tau) \Phi(\tau) + \ApartI(\tau) \Phi^{(1)}(\tau)
\right)
\\
A^{(2)}({\tau}) &=
c \left(
\ApartI^{(2)}(\tau) \Phi(\tau) + 2 \ApartI^{(1)}(\tau) \Phi^{(1)}(\tau) + \ApartI(\tau) \Phi^{(2)}(\tau)
\right)
\,.
\end{align}
\end{subequations}
We have given the components necessary to obtain $\Phi^{(n)}(\tau)$ in the previous section; let us now show how to obtain $\ApartI^{(n)}(\tau)$.
We start by presenting some derivatives that will be useful later.
Firstly, we use Jacobi's formula and its derivative\cite{MatrixCookbook}%
\begin{subequations}
\begin{align}
\label{eq:Jacobi1}\frac{\d}{\d \tau} \det \ApartII &= \ApartII_1 \det \ApartII
\\
\label{eq:Jacobi2} 
\frac{\d^2}{\d \tau^2} \det \ApartII &= \ApartII_2 \det \ApartII
\,,
\end{align}
with
\begin{align}
\ApartII_1 &= \tr \left( \ApartII^{-1} \frac{\d }{\d \tau} \ApartII \right)
\\
\ApartII_2 &= 
\left[
\tr( \ApartII^{-1} \frac{\d}{\d \tau}\ApartII )
\right]^2
+
\tr \left(
\ApartII^{-1}
\frac{\d^2}{\d \tau^2}\ApartII
\right)
-
\tr\left(\left[
\ApartII^{-1}
\left(
\frac{\d}{\d \tau}\ApartII
\right)\right]^2
\right).
\end{align}
\end{subequations}
All the time derivatives of $\ApartII$ can be evaluated analytically [cf.\ Appendix~\ref{app:total_t}] and the inverse of $\ApartII$ can be computed numerically.

Building on these results, we can now calculate the derivatives of the expression $\dta^{N_af}\dtb^{N_bf}\det \ApartII$.
\begin{subequations}
\begin{align}
\label{eq:den1}
\frac{\d}{\d \tau}
\left(
\dta^{N_af}\dtb^{N_bf}
\det \ApartII
\right) 
&=
D_1\, \dta^{N_af}\dtb^{N_bf}
\det \ApartII
\\
\label{eq:den2}
\frac{\d^2}{\d \tau^2}
\left(
\dta^{N_af}\dtb^{N_bf}
\det \ApartII
\right)
&=
D_2\, \dta^{N_af}\dtb^{N_bf}
\det \ApartII 
\,,
\end{align}
with
\begin{align}
D_1
&=
\left( 
\frac{f}{\dta} - \frac{f}{\dtb}
\right)
+
\ApartII_1
\\\notag
D_2 &=
\Bigg\{
\left[
\left( \frac{f}{\dta^2} \frac{N_af-1}{N_a}\right)
-
2
\frac{f}{\dta}\frac{f}{\dtb}
+
\left( \frac{f}{\dtb^2} \frac{N_bf-1}{N_b}\right)
\right]
\\&\phantom{=}+
\left(
\frac{f}{\dta} - \frac{f}{\dtb}
\right)
\ApartII_1
+
\ApartII_2
\Bigg\}
\,.
\end{align}
\end{subequations}

We can now return to $\ApartI(\tau)$ and Taylor expand around $\tau$:
\begin{align}
\nonumber
\ApartI (\tau+h)
&\sim
\ApartI(\tau)
\sqrt{ \frac{1}{ 1 + D_1 h + \frac{1}{2}D_2 h^2  + \mathcal{O}(h^3) }}
\\&\sim
\ApartI(\tau) 
\left(
1 - \frac{1}{2}D_1 h - \frac{1}{8}(2D_2-3D_1^2) h^2 + \mathcal{O}(h^3)
\right)
\,.
\end{align}
Thus, we can deduce that
\begin{subequations}
\begin{align}
\label{eq:f1}
\frac{\d}{\d \tau} \ApartI(\tau) &= -\frac{1}{2}D_1 \times \ApartI(\tau)
\\
\label{eq:f2}
\frac{\d}{\d \tau^2} \ApartI(\tau) &= - \frac{1}{8}(2D_2-3D_1^2) \times \ApartI(\tau)
\,.
\end{align}
\end{subequations}
And the real-time derivatives $\ApartI^{(n)}(\tau)$ necessary for Eq.~\eqref{eq:An} can be obtained using the Cauchy--Riemann condition [Eq.~\eqref{eq:Cauchy_Riemann}].

\section{Analytical results for the Eckart barrier}
\label{app:Eckart}

In this Appendix, we show how to analytically derive the first-order corrected rate for the symmetric Eckart barrier defined in Eq.~\eqref{eq:Eckart_sym}.
We do so using the exact formula for the transmission probability\cite{Eckart} as it depends on $\hbar$
\begin{align}
\label{eq:EckTrans}
P(E) &= \frac{\sinh^2 \left(\EckParam \sqrt{E/V_0} / \hbar \right) }{\sinh^2 \left(\EckParam \sqrt{E/V_0} / \hbar \right) + \cosh^2 \left(\sqrt{\EckParam^2/\hbar^2 - \pi^2 / 4}\right)}
\,,
\end{align}
with $\EckParam = \alpha \hbar = \pi \sqrt{2ma^2 V_0}$.
We can then obtain the exact rate by thermally integrating this probability,
\begin{align}
\label{eq:ThermalRate}
k  &= \frac{1}{\Zr}\frac{1}{2\pi\hbar}\int_0^{\infty} P(E) \, \e^{-\beta E} \, \d E \,,
\end{align}
with
\begin{align}
\Zr &= \sqrt{\frac{m}{2\pi\beta\hbar^2}} \,.
\end{align}

Coming back to instanton theory, we will use neither flux--flux correlation functions nor discretized path integrals to derive our formulas. Yet, we expect to arrive at the same asymptotic expression from Eq.~\eqref{eq:k_asymptotic} as we do when using the general theory developed in the rest of the paper---this is guaranteed by the uniqueness of asymptotic series.

Our first step will be to find an asymptotic expression for $P(E)$ as $\hbar\to 0$.
In particular, we go beyond the lowest-order expansion. %
Let us focus on the components of Eq.~\eqref{eq:EckTrans}.
We begin by noting that
\begin{align}
\label{eq:Eck_sinh}
\sinh^2 \left(\EckParam \sqrt{E/V_0} /\hbar\right)
    &\sim \tfrac{1}{4}\,\e^{2 \EckParam \sqrt{E/V_0} /\hbar}
\,,
\end{align}
to which the correction terms are exponentially small as $\hbar\to 0$.
Next we note that
\begin{align}
\sqrt{\EckParam^2/\hbar^2 - \pi^2 / 4} \sim \frac{1}{\hbar}\EckParam  - \hbar \frac{\pi^2}{8\EckParam}
+
\mathcal{O}(\hbar^2)
\,.
\end{align}
Therefore,
\begin{align}
\label{eq:Eck_cosh}
\cosh (\sqrt{\EckParam^2/\hbar^2 - \pi^2 / 4}) &\sim
\frac{1}{2}
\,\e^{\EckParam / \hbar}  \left(1 - \hbar \frac{\pi^2}{8\EckParam} 
+\mathcal{O}(\hbar^2)
\right) \,.
\end{align}

With this, we can return to Eq.~\eqref{eq:EckTrans} and plug in Eqs.~\eqref{eq:Eck_cosh} and \eqref{eq:Eck_sinh} to obtain:
\begin{align}
\label{eq:Eck_P1}
P(E) \sim
P_1(E) = 
\frac{1}{1 + \e^{W(E)/\hbar}\left(1 - \hbar \frac{\pi^2}{4\EckParam} \right) }
\,,
\end{align}
with
\begin{align}
\label{eq:WEckart_sym}
W(E) = 2\EckParam \left( 1 - \sqrt{E/V_0} \right) \,,    
\end{align}
which is equivalent to the abbreviated action of the instanton path.

It has been noted\cite{PollakRatehbar2,Pollak2022hbarSquared_Correction} that the leading-order expression $P_0(E) =[1 + \exp(W(E)/\hbar) ]^{-1}$ incorrectly predicts that the transmission probability is equal to a half when the energy equals the barrier height, $P_0(V_0)=1/2$. 
We observe that the present first-order correction also remedies this behavior. In particular, the expression $P_1(V_0)$ for the symmetric Eckart barrier is greater than $1/2$. As we shall see in the following paragraph, this is a key component of the correction in RPI+PC theory, and hence the first order RPI+PC is entirely consistent with the ``$\hbar^2$'' correction of Refs.~\onlinecite{PollakRatehbar2} and \citenum{Pollak2022hbarSquared_Correction}.%

To evaluate the rate constant asymptotically as $\hbar\to0$ for fixed $\tot$, we first rewrite Eq.~\eqref{eq:ThermalRate} to make all the $\hbar$ and $\tot$ dependence explicit,
\begin{align}
\label{eq:ThermalRate2}
k  &= \sqrt{\frac{\tot}{2\pi m}}\int_0^{\infty} P(E;\hbar) \, \e^{-\tot E/\hbar} \, \d E \,.
\end{align}
Now to evaluate this integral asymptotically we need to consider the behavior of the integrand as $\hbar\to 0$. Using Eq.~\eqref{eq:Eck_P1} in the deep-tunneling regime, where $\tot<2\pi /\omega$ (and $\omega=\sqrt{2V_0/(ma^2)}$ is the barrier frequency), one finds that as $\hbar\to0$ the integral is dominated by the region $0<E<V_0$. It is then straightforward to show that in this region the integrand can be approximated asymptotically as  
\begin{equation}
    P(E;\hbar)\,\e^{-\tot E/\hbar} \sim \e^{-W(E)/\hbar-\tot E/\hbar}\left(1+\hbar \frac{\pi^2}{4\gamma} + \mathcal{O}(\hbar^2)\right).
\end{equation}
Hence, integrating using the first-order steepest-descent method described in Sec.~\ref{sec:cff}, we expand around the stationary energy $\tilde{E} = V_0 {\EckParam^2}/{(\tot)^2}$,
to obtain
\begin{subequations}
\begin{align}
\label{eq:Eck_res}
k &\sim
k_0
\big[
1
+
\hbar(\Gamma_{\mathrm A}+\Gamma_{\mathrm B}+\Gamma_{\mathrm C})
+\mathcal{O}(\hbar^2)\big]
\\
k_0 &= \sqrt{\frac{\tot}{2\pi m}}
\, \e^{-\left[W(\tilde{E}) +\tot \tilde{E}\right]/\hbar}
\sqrt{\frac{2\pi\hbar}{W^{(2)}}}
\\
\Gamma_{\mathrm A} &= -\frac{3}{4!}  \frac{W^{(4)}}{[W^{(2)}]^2}
\\
\Gamma_{\mathrm B} &= \frac{15}{2\times(3!)^2} \frac{[W^{(3)}]^2}{[W^{(2)}]^3}
\\
\Gamma_{\mathrm C} &=  \frac{\pi^2}{4\EckParam}
\,,
\end{align}
\end{subequations}
where $W^{(n)}$ is the $n$-th derivative of $W(E)$ evaluated at the stationary point $\tilde{E}$.
Note that in general, the coefficients $\Gamma_{\mathrm A}$, $\Gamma_{\mathrm B}$ and $\Gamma_{\mathrm C}$ do not directly correspond to the $\Gammax$ and $\Gammat$ coefficients of the RPI theory---only their sum $\Gamma_1$ is the same for both theories (in the limit $N\to\infty$).

Using Eq.~\eqref{eq:WEckart_sym} to evaluate the abbreviated action, we notice that
$\Gamma_{\mathrm A} = -\Gamma_{\mathrm B} = -15 \sqrt{V_0/\tilde{E}}/(16\EckParam) $.
Thus, only the temperature-independent $\Gamma_{\mathrm C}$ contributes to the first-order correction for the symmetric Eckart barrier.

\bibliography{extra, references}

\end{document}


\title{Supplementary Information: Perturbatively corrected ring-polymer instanton rate theory rigorously captures anharmonicity and deep tunneling}
\author{Jindra Du\v sek}
\affiliation{Department of Chemistry and Applied Biosciences, ETH Z\"urich, 8093 Z\"urich, Switzerland}
\author{Joseph E. Lawrence}
\affiliation{Department of Chemistry and Applied Biosciences, ETH Z\"urich, 8093 Z\"urich, Switzerland}
\affiliation{\mbox{Simons Center for Computational Physical Chemistry, New York University, New York, NY 10003, USA}}
\affiliation{Department of Chemistry, New York University, New York, NY 10003, USA}
\author{Jeremy O. Richardson}
\email{jeremy.richardson@phys.chem.ethz.ch}
\affiliation{Department of Chemistry and Applied Biosciences, ETH Z\"urich, 8093 Z\"urich, Switzerland}
\date{\today}

\maketitle

\onecolumngrid

\section{$N\to\infty$ convergence for the symmetric Eckart barrier}
\label{sec:Nconvergence}

\begin{figure}[b]
    \centering
    \includegraphics[width=.5\textwidth]{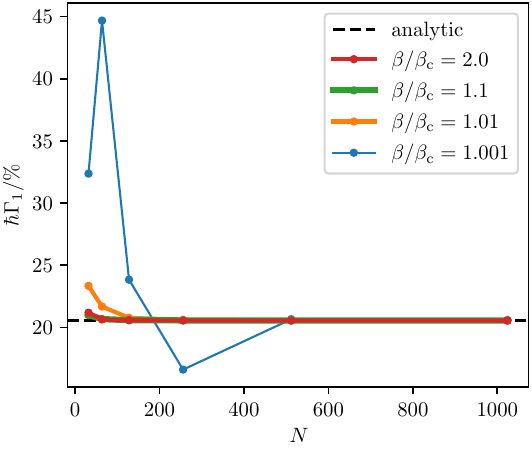}
    \caption{The first-order correction $\hbar\Gamma_1$ computed at different temperatures. Each plot shows the convergence with the number of beads $N$.
    Four different values of $\beta$ were chosen to approach the crossover temperature.
    }
    \label{fig:EckConvBeta2}
\end{figure}

\begin{figure}[b]
    \centering
    \includegraphics[width=\textwidth]{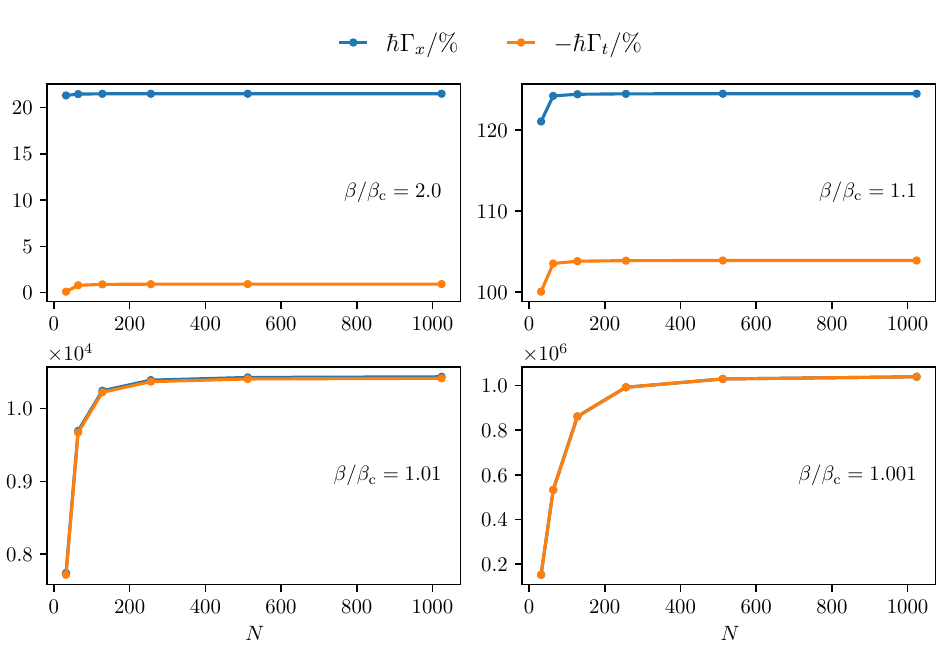}
    \caption{The first-order correction components $\hbar\Gammax$ and $-\hbar\Gammat$ computed at different temperatures. Each plot shows the convergence with the number of beads $N$. The first-order correction $\hbar\Gamma_1=\hbar\Gammax+\hbar\Gammat$ can be obtained as a difference of the two plotted curves.
    Four different values of $\beta$ were chosen to approach the crossover temperature.
    }
    \label{fig:EckConvBeta}
\end{figure}

In Appendix C of the main paper, we prove that for the symmetric Eckart barrier, the first-order correction does not depend on temperature and is equal to $\hbar\Gamma_1 = \hbar\pi^2/(4\gamma)=20.6\%$.
We have obtained the same result using discretized path integrals and in Fig.~\ref{fig:EckConvBeta2}, we show the convergence of $\hbar \Gamma_1$ with the number of beads $N$.
It is notable, however, that the convergence slows down for instantons %
near the crossover temperature.
We study this in more detail in Fig.~\ref{fig:EckConvBeta}, where the individual convergence of the components $\hbar\Gammax$ and $-\hbar\Gammat$ is shown. Although $\hbar\Gammax$ and $-\hbar\Gammat$ do converge with the number of beads for a fixed $\beta$, their magnitudes increase as the crossover temperature is approached.
However, the cancellation between the two contributions recovers the correct result for the total $\hbar\Gamma_1$ term in the $N\to\infty$ limit.

\clearpage

\section{Collinear \HHtwo reaction data}

We present the numerical values of the calculated rates plotted in Fig.~6 of the main paper in Tables~\ref{table:HpH2}, \ref{table:DpD2} and \ref{table:TpT2}.

\begin{table}[b]
\caption{Rates and their relative errors computed for the \HHtwo reaction. Instanton rates were computed with $N=1024$ beads.
All rates are given in $\mathrm{cm} /(\mathrm{molecule} \times s)$.
The exact rates below 110 K are converged to $\pm2$ units in the third significant figure.}
\label{table:HpH2}
\begin{tabular}{c cccc cccc}
\toprule
\head{$T/\mathrm{K}$} &
\head{$k_{\mathrm{exact}}$}  &
\head{$k_{\mathrm{inst}, 0}$} &
\head{$k_{\mathrm{inst}, 1}$} &
\head{$k_{\mathrm{inst}, 1\mathrm{c}}$} &
\head{$\Gamma_1 / \%$}  & 
\head{$\delta_0 / \%$}  & 
\head{$\delta_1 / \%$}  &
\head{$\delta_{1\mathrm{c}} / \%$}
\\
\midrule
 75 & $5.87\!\times\! 10^{-6}$ & $4.118\!\times\! 10^{-6}$ & $5.206\!\times\! 10^{-6}$ & $5.364\!\times\! 10^{-6}$ & $26$ & $-30$ & $-11$ & $-8.6$  \\
 80 & $8.95\!\times\! 10^{-6}$ & $6.238\!\times\! 10^{-6}$ & $7.893\!\times\! 10^{-6}$ & $8.134\!\times\! 10^{-6}$ & $27$ & $-30$ & $-12$ & $-9.1$  \\
 85 & $1.36\!\times\! 10^{-5}$ & $9.444\!\times\! 10^{-6}$ & $1.197\!\times\! 10^{-5}$ & $1.234\!\times\! 10^{-5}$ & $27$ & $-31$ & $-12$ & $-9.3$  \\
 90 & $2.08\!\times\! 10^{-5}$ & $1.428\!\times\! 10^{-5}$ & $1.815\!\times\! 10^{-5}$ & $1.873\!\times\! 10^{-5}$ & $27$ & $-31$ & $-13$ & $-10$  \\
 95 & $3.18\!\times\! 10^{-5}$ & $2.155\!\times\! 10^{-5}$ & $2.749\!\times\! 10^{-5}$ & $2.839\!\times\! 10^{-5}$ & $28$ & $-32$ & $-14$ & $-11$  \\
100 & $4.85\!\times\! 10^{-5}$ & $3.246\!\times\! 10^{-5}$ & $4.156\!\times\! 10^{-5}$ & $4.297\!\times\! 10^{-5}$ & $28$ & $-33$ & $-14$ & $-11$  \\
110 & $1.12\!\times\! 10^{-4}$ & $7.300\!\times\! 10^{-5}$ & $9.440\!\times\! 10^{-5}$ & $9.787\!\times\! 10^{-5}$ & $29$ & $-35$ & $-16$ & $-13$  \\
120 & $2.59\!\times\! 10^{-4}$ & $1.621\!\times\! 10^{-4}$ & $2.121\!\times\! 10^{-4}$ & $2.206\!\times\! 10^{-4}$ & $31$ & $-37$ & $-18$ & $-15$  \\
130 & $5.90\!\times\! 10^{-4}$ & $3.548\!\times\! 10^{-4}$ & $4.704\!\times\! 10^{-4}$ & $4.915\!\times\! 10^{-4}$ & $33$ & $-40$ & $-20$ & $-17$  \\
140 & $1.32\!\times\! 10^{-3}$ & $7.645\!\times\! 10^{-4}$ & $1.029\!\times\! 10^{-3}$ & $1.080\!\times\! 10^{-3}$ & $35$ & $-42$ & $-22$ & $-18$  \\
150 & $2.89\!\times\! 10^{-3}$ & $1.621\!\times\! 10^{-3}$ & $2.216\!\times\! 10^{-3}$ & $2.340\!\times\! 10^{-3}$ & $37$ & $-44$ & $-23$ & $-19$  \\
160 & $6.14\!\times\! 10^{-3}$ & $3.379\!\times\! 10^{-3}$ & $4.700\!\times\! 10^{-3}$ & $4.996\!\times\! 10^{-3}$ & $39$ & $-45$ & $-23$ & $-19$  \\
170 & $1.26\!\times\! 10^{-2}$ & $6.926\!\times\! 10^{-3}$ & $9.807\!\times\! 10^{-3}$ & $1.050\!\times\! 10^{-2}$ & $42$ & $-45$ & $-22$ & $-17$  \\
$\Tc/2\approx 172$ & $1.44\!\times\! 10^{-2}$ & $7.937\!\times\! 10^{-3}$ & $1.128\!\times\! 10^{-2}$ & $1.209\!\times\! 10^{-2}$ & $42$ & $-45$ & $-22$ & $-16$  \\
180 & $2.48\!\times\! 10^{-2}$ & $1.395\!\times\! 10^{-2}$ & $2.012\!\times\! 10^{-2}$ & $2.172\!\times\! 10^{-2}$ & $44$ & $-44$ & $-19$ & $-12$  \\
190 & $4.70\!\times\! 10^{-2}$ & $2.760\!\times\! 10^{-2}$ & $4.059\!\times\! 10^{-2}$ & $4.419\!\times\! 10^{-2}$ & $47$ & $-41$ & $-14$ & $-6$  \\
200 & $8.56\!\times\! 10^{-2}$ & $5.364\!\times\! 10^{-2}$ & $8.044\!\times\! 10^{-2}$ & $8.840\!\times\! 10^{-2}$ & $50$ & $-37$ & $-6$ & $3.3$  \\
210 & $1.50\!\times\! 10^{-1}$ & $1.024\!\times\! 10^{-1}$ & $1.565\!\times\! 10^{-1}$ & $1.737\!\times\! 10^{-1}$ & $53$ & $-32$ & $4.3$ & $16$  \\
220 & $2.54\!\times\! 10^{-1}$ & $1.919\!\times\! 10^{-1}$ & $2.989\!\times\! 10^{-1}$ & $3.351\!\times\! 10^{-1}$ & $56$ & $-24$ & $18$ & $32$  \\
230 & $4.15\!\times\! 10^{-1}$ & $3.530\!\times\! 10^{-1}$ & $5.594\!\times\! 10^{-1}$ & $6.334\!\times\! 10^{-1}$ & $58$ & $-15$ & $35$ & $53$  \\
240 & $6.57\!\times\! 10^{-1}$ & $6.372\!\times\! 10^{-1}$ & $1.025\!\times\! 10^{0}$ & $1.172\!\times\! 10^{0}$ & $61$ & $-3$ & $56$ & $78$  \\
250 & $1.01\!\times\! 10^{0}$ & $1.128\!\times\! 10^{0}$ & $1.837\!\times\! 10^{0}$ & $2.115\!\times\! 10^{0}$ & $63$ & $12$ & $82$ & $1.1\!\times\! 10^{2}$  \\
260 & $1.51\!\times\! 10^{0}$ & $1.957\!\times\! 10^{0}$ & $3.211\!\times\! 10^{0}$ & $3.715\!\times\! 10^{0}$ & $64$ & $30$ & $1.1\!\times\! 10^{2}$ & $1.5\!\times\! 10^{2}$  \\
270 & $2.21\!\times\! 10^{0}$ & $3.326\!\times\! 10^{0}$ & $5.465\!\times\! 10^{0}$ & $6.327\!\times\! 10^{0}$ & $64$ & $50$ & $1.5\!\times\! 10^{2}$ & $1.9\!\times\! 10^{2}$  \\
280 & $3.16\!\times\! 10^{0}$ & $5.535\!\times\! 10^{0}$ & $9.020\!\times\! 10^{0}$ & $1.039\!\times\! 10^{1}$ & $63$ & $75$ & $1.9\!\times\! 10^{2}$ & $2.3\!\times\! 10^{2}$  \\
290 & $4.42\!\times\! 10^{0}$ & $9.011\!\times\! 10^{0}$ & $1.437\!\times\! 10^{1}$ & $1.633\!\times\! 10^{1}$ & $59$ & $1.0\!\times\! 10^{2}$ & $2.3\!\times\! 10^{2}$ & $2.7\!\times\! 10^{2}$  \\
300 & $6.06\!\times\! 10^{0}$ & $1.435\!\times\! 10^{1}$ & $2.186\!\times\! 10^{1}$ & $2.421\!\times\! 10^{1}$ & $52$ & $1.4\!\times\! 10^{2}$ & $2.6\!\times\! 10^{2}$ & $3.0\!\times\! 10^{2}$  \\
310 & $8.16\!\times\! 10^{0}$ & $2.235\!\times\! 10^{1}$ & $3.087\!\times\! 10^{1}$ & $3.272\!\times\! 10^{1}$ & $38$ & $1.7\!\times\! 10^{2}$ & $2.8\!\times\! 10^{2}$ & $3.0\!\times\! 10^{2}$  \\
320 & $1.08\!\times\! 10^{1}$ & $3.406\!\times\! 10^{1}$ & $3.589\!\times\! 10^{1}$ & $3.594\!\times\! 10^{1}$ & $5.4$ & $2.2\!\times\! 10^{2}$ & $2.3\!\times\! 10^{2}$ & $2.3\!\times\! 10^{2}$  \\
330 & $1.41\!\times\! 10^{1}$ & $5.079\!\times\! 10^{1}$ & $-5.306\!\times\! 10^{0}$ & $1.683\!\times\! 10^{1}$ & $-1.1\!\times\! 10^{2}$ & $2.6\!\times\! 10^{2}$ & $-1.4\!\times\! 10^{2}$ & $19$  \\
340 & $1.82\!\times\! 10^{1}$ & $7.422\!\times\! 10^{1}$ & $-1.430\!\times\! 10^{3}$ & $1.176\!\times\! 10^{-7}$ & $-2.0\!\times\! 10^{3}$ & $3.1\!\times\! 10^{2}$ & $-8.0\!\times\! 10^{3}$ & $-1.0\!\times\! 10^{2}$  \\
\bottomrule
\end{tabular}
\end{table}

\begin{table}[t]
\caption{Rates and their relative errors computed for the \DDtwo reaction. Instanton rates were computed with $N=1024$ beads.
All rates are given in $\mathrm{cm} /(\mathrm{molecule} \times s)$.
The exact rates below 110 K are converged to $\pm2$ units in the third significant figure.
}
\label{table:DpD2}
\begin{tabular}{c cccc cccc
}
\toprule
\head{$T/\mathrm{K}$} &
\head{$k_{\mathrm{exact}}$}  &
\head{$k_{\mathrm{inst}, 0}$} &
\head{$k_{\mathrm{inst}, 1}$} &
\head{$k_{\mathrm{inst}, 1\mathrm{c}}$} &
\head{$\Gamma_1 / \%$}  & 
\head{$\delta_0 / \%$}  & 
\head{$\delta_1 / \%$}  &
\head{$\delta_{1\mathrm{c}} / \%$}
\\
\midrule
 75 & $5.40\!\times\! 10^{-10}$ & $4.161\!\times\! 10^{-10}$ & $5.008\!\times\! 10^{-10}$ & $5.100\!\times\! 10^{-10}$ & $20$ & $-23$ & $-7.3$ & $-5.6$  \\
 80 & $1.19\!\times\! 10^{-9}$ & $9.058\!\times\! 10^{-10}$ & $1.096\!\times\! 10^{-9}$ & $1.118\!\times\! 10^{-9}$ & $21$ & $-24$ & $-7.9$ & $-6.1$  \\
 85 & $2.61\!\times\! 10^{-9}$ & $1.957\!\times\! 10^{-9}$ & $2.384\!\times\! 10^{-9}$ & $2.434\!\times\! 10^{-9}$ & $22$ & $-25$ & $-8.7$ & $-6.7$  \\
 90 & $5.67\!\times\! 10^{-9}$ & $4.189\!\times\! 10^{-9}$ & $5.140\!\times\! 10^{-9}$ & $5.257\!\times\! 10^{-9}$ & $23$ & $-26$ & $-9.3$ & $-7.3$  \\
 95 & $1.22\!\times\! 10^{-8}$ & $8.885\!\times\! 10^{-9}$ & $1.099\!\times\! 10^{-8}$ & $1.125\!\times\! 10^{-8}$ & $24$ & $-27$ & $-9.9$ & $-7.8$  \\
100 & $2.62\!\times\! 10^{-8}$ & $1.866\!\times\! 10^{-8}$ & $2.325\!\times\! 10^{-8}$ & $2.387\!\times\! 10^{-8}$ & $25$ & $-29$ & $-11$ & $-8.9$  \\
110 & $1.17\!\times\! 10^{-7}$ & $7.964\!\times\! 10^{-8}$ & $1.010\!\times\! 10^{-7}$ & $1.042\!\times\! 10^{-7}$ & $27$ & $-32$ & $-14$ & $-11$  \\
120 & $4.95\!\times\! 10^{-7}$ & $3.250\!\times\! 10^{-7}$ & $4.205\!\times\! 10^{-7}$ & $4.360\!\times\! 10^{-7}$ & $29$ & $-34$ & $-15$ & $-12$  \\
$\Tc/2\approx 122$ & $6.22\!\times\! 10^{-7}$ & $4.063\!\times\! 10^{-7}$ & $5.274\!\times\! 10^{-7}$ & $5.474\!\times\! 10^{-7}$ & $30$ & $-35$ & $-15$ & $-12$  \\
130 & $1.97\!\times\! 10^{-6}$ & $1.266\!\times\! 10^{-6}$ & $1.672\!\times\! 10^{-6}$ & $1.745\!\times\! 10^{-6}$ & $32$ & $-36$ & $-15$ & $-11$  \\
140 & $7.27\!\times\! 10^{-6}$ & $4.704\!\times\! 10^{-6}$ & $6.346\!\times\! 10^{-6}$ & $6.670\!\times\! 10^{-6}$ & $35$ & $-35$ & $-13$ & $-8.3$  \\
150 & $2.45\!\times\! 10^{-5}$ & $1.665\!\times\! 10^{-5}$ & $2.295\!\times\! 10^{-5}$ & $2.430\!\times\! 10^{-5}$ & $38$ & $-32$ & $-6.3$ & $-0.82$  \\
160 & $7.57\!\times\! 10^{-5}$ & $5.611\!\times\! 10^{-5}$ & $7.891\!\times\! 10^{-5}$ & $8.424\!\times\! 10^{-5}$ & $41$ & $-26$ & $4.2$ & $11$  \\
170 & $2.14\!\times\! 10^{-4}$ & $1.799\!\times\! 10^{-4}$ & $2.574\!\times\! 10^{-4}$ & $2.769\!\times\! 10^{-4}$ & $43$ & $-16$ & $20$ & $29$  \\
180 & $5.54\!\times\! 10^{-4}$ & $5.479\!\times\! 10^{-4}$ & $7.941\!\times\! 10^{-4}$ & $8.587\!\times\! 10^{-4}$ & $45$ & $-1.1$ & $43$ & $55$  \\
190 & $1.33\!\times\! 10^{-3}$ & $1.584\!\times\! 10^{-3}$ & $2.304\!\times\! 10^{-3}$ & $2.496\!\times\! 10^{-3}$ & $46$ & $19$ & $73$ & $88$  \\
200 & $2.98\!\times\! 10^{-3}$ & $4.337\!\times\! 10^{-3}$ & $6.248\!\times\! 10^{-3}$ & $6.738\!\times\! 10^{-3}$ & $44$ & $46$ & $1.1\!\times\! 10^{2}$ & $1.3\!\times\! 10^{2}$  \\
210 & $6.24\!\times\! 10^{-3}$ & $1.124\!\times\! 10^{-2}$ & $1.562\!\times\! 10^{-2}$ & $1.660\!\times\! 10^{-2}$ & $39$ & $80$ & $1.5\!\times\! 10^{2}$ & $1.7\!\times\! 10^{2}$  \\
220 & $1.24\!\times\! 10^{-2}$ & $2.758\!\times\! 10^{-2}$ & $3.461\!\times\! 10^{-2}$ & $3.559\!\times\! 10^{-2}$ & $25$ & $1.2\!\times\! 10^{2}$ & $1.8\!\times\! 10^{2}$ & $1.9\!\times\! 10^{2}$  \\
230 & $2.33\!\times\! 10^{-2}$ & $6.409\!\times\! 10^{-2}$ & $4.930\!\times\! 10^{-2}$ & $5.088\!\times\! 10^{-2}$ & $-23$ & $1.8\!\times\! 10^{2}$ & $1.1\!\times\! 10^{2}$ & $1.2\!\times\! 10^{2}$  \\
\bottomrule
\end{tabular}
\end{table}

\begin{table}[t]
\caption{Rates and their relative errors computed for the \TTtwo reaction. Instanton rates were computed with $N=1024$ beads.
All rates are given in $\mathrm{cm} /(\mathrm{molecule} \times s)$.
The exact rates below 110 K are converged to $\pm2$ units in the third significant figure.
}
\label{table:TpT2}
\begin{tabular}{c cccc cccc
}
\toprule
\head{$T/\mathrm{K}$} &
\head{$k_{\mathrm{exact}}$}  &
\head{$k_{\mathrm{inst}, 0}$} &
\head{$k_{\mathrm{inst}, 1}$} &
\head{$k_{\mathrm{inst}, 1\mathrm{c}}$} &
\head{$\Gamma_1 / \%$}  & 
\head{$\delta_0 / \%$}  & 
\head{$\delta_1 / \%$}  &
\head{$\delta_{1\mathrm{c}} / \%$}
\\
\midrule
 75 & $1.58\!\times\! 10^{-12}$ & $1.243\!\times\! 10^{-12}$ & $1.477\!\times\! 10^{-12}$ & $1.500\!\times\! 10^{-12}$ & $19$ & $-21$ & $-6.5$ & $-5.1$  \\
 80 & $4.84\!\times\! 10^{-12}$ & $3.751\!\times\! 10^{-12}$ & $4.494\!\times\! 10^{-12}$ & $4.572\!\times\! 10^{-12}$ & $20$ & $-23$ & $-7.1$ & $-5.5$  \\
 85 & $1.46\!\times\! 10^{-11}$ & $1.112\!\times\! 10^{-11}$ & $1.344\!\times\! 10^{-11}$ & $1.370\!\times\! 10^{-11}$ & $21$ & $-24$ & $-7.9$ & $-6.2$  \\
 90 & $4.32\!\times\! 10^{-11}$ & $3.232\!\times\! 10^{-11}$ & $3.943\!\times\! 10^{-11}$ & $4.027\!\times\! 10^{-11}$ & $22$ & $-25$ & $-8.7$ & $-6.8$  \\
 95 & $1.26\!\times\! 10^{-10}$ & $9.211\!\times\! 10^{-11}$ & $1.135\!\times\! 10^{-10}$ & $1.162\!\times\! 10^{-10}$ & $23$ & $-27$ & $-9.9$ & $-7.8$  \\
$\Tc/2\approx 99$ & $3.16\!\times\! 10^{-10}$ & $2.269\!\times\! 10^{-10}$ & $2.821\!\times\! 10^{-10}$ & $2.895\!\times\! 10^{-10}$ & $24$ & $-28$ & $-11$ & $-8.4$  \\
100 & $3.59\!\times\! 10^{-10}$ & $2.572\!\times\! 10^{-10}$ & $3.202\!\times\! 10^{-10}$ & $3.286\!\times\! 10^{-10}$ & $25$ & $-28$ & $-11$ & $-8.5$  \\
110 & $2.72\!\times\! 10^{-9}$ & $1.882\!\times\! 10^{-9}$ & $2.395\!\times\! 10^{-9}$ & $2.472\!\times\! 10^{-9}$ & $27$ & $-31$ & $-12$ & $-9.1$  \\
120 & $1.82\!\times\! 10^{-8}$ & $1.264\!\times\! 10^{-8}$ & $1.645\!\times\! 10^{-8}$ & $1.709\!\times\! 10^{-8}$ & $30$ & $-31$ & $-9.6$ & $-6.1$  \\
130 & $1.06\!\times\! 10^{-7}$ & $7.779\!\times\! 10^{-8}$ & $1.035\!\times\! 10^{-7}$ & $1.082\!\times\! 10^{-7}$ & $33$ & $-27$ & $-2.4$ & $2.1$  \\
140 & $5.35\!\times\! 10^{-7}$ & $4.381\!\times\! 10^{-7}$ & $5.935\!\times\! 10^{-7}$ & $6.247\!\times\! 10^{-7}$ & $35$ & $-18$ & $11$ & $17$  \\
150 & $2.32\!\times\! 10^{-6}$ & $2.254\!\times\! 10^{-6}$ & $3.088\!\times\! 10^{-6}$ & $3.264\!\times\! 10^{-6}$ & $37$ & $-2.8$ & $33$ & $41$  \\
160 & $8.74\!\times\! 10^{-6}$ & $1.057\!\times\! 10^{-5}$ & $1.445\!\times\! 10^{-5}$ & $1.526\!\times\! 10^{-5}$ & $37$ & $21$ & $65$ & $75$  \\
170 & $2.91\!\times\! 10^{-5}$ & $4.506\!\times\! 10^{-5}$ & $5.993\!\times\! 10^{-5}$ & $6.268\!\times\! 10^{-5}$ & $33$ & $55$ & $1.1\!\times\! 10^{2}$ & $1.2\!\times\! 10^{2}$  \\
180 & $8.66\!\times\! 10^{-5}$ & $1.747\!\times\! 10^{-4}$ & $2.102\!\times\! 10^{-4}$ & $2.141\!\times\! 10^{-4}$ & $20$ & $1.0\!\times\! 10^{2}$ & $1.4\!\times\! 10^{2}$ & $1.5\!\times\! 10^{2}$  \\
190 & $2.34\!\times\! 10^{-4}$ & $6.171\!\times\! 10^{-4}$ & $3.238\!\times\! 10^{-4}$ & $3.836\!\times\! 10^{-4}$ & $-48$ & $1.6\!\times\! 10^{2}$ & $38$ & $64$  \\
\bottomrule
\end{tabular}
\end{table}

\bibliography{extra, references}